\documentclass{aa}

\usepackage{graphicx}

\usepackage{txfonts}
\usepackage{xcolor}
\usepackage{lipsum}
\usepackage{subcaption} 
\usepackage{lscape}             
\usepackage{placeins}           
\usepackage[colorlinks=true, linkcolor=blue, citecolor=blue, filecolor=blue, urlcolor=blue]{hyperref}

\newcommand{\gaia}{{\it Gaia}\xspace}
\newcommand{\kima}{{\tt kima}\xspace}
\newcommand{\np}{\ensuremath{N_{\rm p}}\xspace}
\newcommand{\eq}[1]{Equation \ref{#1}\xspace}
\newcommand{\tbl}[1]{Table \ref{#1}\xspace}
\newcommand{\fig}[1]{Figure \ref{#1}\xspace}
\newcommand{\sect}[1]{Section \ref{#1}\xspace}

\newcommand{\edit}[1]{#1}
\newcommand{\editt}[1]{#1}

\newcommand{\gaiamodel}{{\tt GAIAmodel}\xspace}
\newcommand{\rvgaiamodel}{{\tt RVGAIAmodel}\xspace}
\newcommand{\kepmodel}{{\tt kepmodel}\xspace}
\begin{document}

   \title{Bayesian analysis of Gaia epoch astrometry and radial velocities with \kima}

   \subtitle{}

   \author{
        T. A. Baycroft\inst{1,2}\fnmsep\thanks{Corresponding author: thomasbaycroftastro@gmail.com}
        \and J. P. Faria \inst{3}
        \and J-B. Delisle \inst{3}
    }

   \institute{Tsung-Dao Lee Institute, Shanghai Jiao Tong University, 1 Lisuo Road, Shanghai 201210, China
        \and School of Physics and Astronomy, University of Birmingham, Edgbaston, Birmingham B15 2TT, UK
        \and Observatoire Astronomique de l'Université de Genève, Chemin Pegasi 51b, 1290 Versoix, Switzerland \label{ inst3 }
   }

   \date{Accepted September 2026}

  \abstract{
  The forthcoming data release\edit{s} from the \gaia space telescope \edit{are} expected to yield \edit{thousands to} tens of thousands of newly detected exoplanets, as well as detection and 3D orbital constraints on binary stars and black holes. Many of these systems will warrant in-depth analyses of the astrometric data and radial velocity follow-up. This will require dedicated tools to exploit this wealth of data. We provide open-source software to analyse epoch astrometric data from \gaia, both independently and jointly with radial velocities. We add two models to the open-source orbit-fitting codebase \kima and test these on both real and simulated data. This code uses diffusive nested sampling to explore the parameter space, calculate evidence for model comparison, and perform parameter estimation. We show that the results are consistent with published and expected values, validating the use of \kima for the analysis of \gaia data. We explore various attributes of \kima's \gaia model, including its potential to distinguish genuine orbital signals from scan-angle-dependent signals.}


   \maketitle
   \nolinenumbers

\section{Introduction}

The fourth data release (DR4) from the \gaia mission
\citep{gaia_collaboration_gaia_2016} is expected in December 2026%
\footnote{See \href{https://www.cosmos.esa.int/web/gaia/data-release-4}{www.cosmos.esa.int/web/gaia/data-release-4}} %
and will include individual epoch astrometry for approximately two billion astrophysical
sources. This rich dataset will enable the orbital characterisation of many
types of objects, including exoplanets \citep[e.g.][]{perryman_astrometric_2014,
stefansson_gaia-4b_2025, lammers_exoplanet_2026}, binary stars
\citep[e.g.][]{gaia_collaboration_gaia_2023-1, halbwachs_gaia_2023,
baycroft_predicting_2026}, and black holes \citep[e.g.][]{breivik_revealing_2017,
el-badry_sun-like_2023, gaia_collaboration_discovery_2024,
nagarajan_realistic_2025}.

The \kima codebase for orbital fitting \citep{faria_kima_2018} was originally developed to analyse radial-velocity observations of CoRoT-7 \citep{faria_uncovering_2016}. The code has subsequently been used in various studies \citep[e.g.][]{faria_decoding_2020, lillo-box_planetary_2020, frensch_harps_2023, john_sub-m_2023} and extended to analyse binary stars \citep{baycroft_improving_2023} and eclipse timing variations \citep{baycroft_new_2023}. The \kima code uses the diffusive nested sampling algorithm \citep{brewer_diffusive_2011} as implemented in the DNEST4  package \citep{brewer_dnest4_2018}. This nested sampling implementation allows, in particular, for trans-dimensional sampling, where the number of planets or Keplerians in the model, denoted as \np, is a free parameter \citep{brewer_fast_2015}. By default, \kima assigns a uniform prior to \np, up to a specified maximum, \(N_{p,{\rm max}}\), and fits all orbital parameters simultaneously, using the same prior distributions. 

Because the number of Keplerians is a free parameter, \kima can efficiently perform Bayesian model
comparison by comparing models with \(\np=n\) and
\(\np=n-1\), as well as  parameter estimation of the detected signals, all
within a single run of \kima \citep[see e.g.][]{faria_candidate_2022}. Moreover, from
this single run (provided \(N_{p, {\rm max}}\) exceeds the number of
significantly detected Keplerian signals), the resulting posterior distribution
for additional signals compatible with the data can be used to generate
`compatibility limits' similar to traditional `detection limits'
\citep[e.g.][]{standing_bebop_2022}. Posterior distributions from the
analysis of several stars can be used to compute Bayesian estimates of planet
occurrence rates \citep{faria_inferring_2025}.

In this work, we present the addition of bespoke models within the \kima
framework to analyse \gaia epoch astrometry alone or jointly with radial
velocities. Previous studies have demonstrated the power of combining astrometry with radial velocities. In particular, combining with the \textit{Hipparcos-Gaia} proper-motion anomaly
method \citep[e.g.][]{brandt_hipparcos-gaia_2021, kervella_stellar_2022}
provides 3D constraints on long-period orbits
\citep[e.g.][]{feng_lessons_2025,wu_detection_2026,piccinini_true_2026,pena_cold_2026}. A model already exists
in \kima for the analysis of proper-motion anomalies, which will be presented in
detail in \citetext{Ceva et al. in prep.}. Other tools can fit \gaia astrometry, such as \edit{\texttt{orvara} \citep{brandt_orvara_2021},} \kepmodel
\citep{delisle_analytical_2022}, \texttt{BINARYS}
\citep{leclerc_combining_2023}, \texttt{Octofitter}
\citep{thompson_octofitter_2023}\edit{, or \texttt{gaiamock} \citep{el-badry_generative_2024}}. 
The efficient and robust nested-sampling algorithm, together with the ability to simultaneously fit any number of Keplerians in a single analysis, distinguishes \kima from these tools. These features make
\kima an efficient tool that can perform model
comparisons for planet detection, parameter estimation of the detected planets, and
subsequent calculations of compatibility limits and occurrence rates all from a single analysis
\citep{faria_inferring_2025}.

The paper is structured as follows. We present the different models  in
\sect{sec:models}, perform tests and applications in \sect{sec:testing}, show
the calculation of compatibility limits in \sect{sec:compat}, and
conclude in \sect{sec:conclusions}.

\section{The models}\label{sec:models}

This section presents two models implemented in \kima: the \gaiamodel, which analyses \gaia epoch astrometry alone, and the\rvgaiamodel, which jointly analyses \gaia epoch astrometry and radial velocities. We also briefly mention the \texttt{RVHGPMmodel}, which is presented in \citetext{Ceva et al. in prep.}. The complete documentation is available at \href{https://www.kima.science}{kima.science}.

\subsection{\gaiamodel}

For \gaia epoch astrometry, the measured variable corresponds to the position in
the along-scan direction relative to the reference position, which we denote
by the letter \(w\). We denote the observed data by \(w_{\rm obs}\) and use
\(w_{\rm X}\), with various subscripts, for the different components of
the modelled position. The equations for the baseline and Keplerian models
follow standard forms \citep[e.g.][]{hilditch_introduction_2001,
sahlmann_search_2011, perryman_astrometric_2014,
perryman_astrometry_2018,ranalli_astrometry_2018,holl_gaia_2023}. The baseline
model is the standard five-parameter single-star model for a scanning telescope,
\begin{equation}\label{eq:baseline}
    w_{\rm ss} = (\Delta\alpha^{\star} + \mu_{\alpha^{\star}}t)\sin{\psi} 
                 + (\Delta\delta + \mu_{\delta}t)\cos{\psi} + \varpi f_{\varpi} ,
\end{equation}
where $\Delta\alpha^{\star}$ and $\Delta\delta$ are offsets from the reference
location, $\alpha_0,\delta_0$, in equatorial coordinates, with
$\Delta\alpha^{\star} = \Delta\alpha\cos{\delta}$. In Eq.~\eqref{eq:baseline},
$\psi$ is the \gaia scan angle at each measurement epoch, defined relative to
the local north%
\footnote{See \href{https://www.cosmos.esa.int/web/gaia/scanning-law-pointings}{www.cosmos.esa.int/web/gaia/scanning-law-pointings}.}; %
$\varpi$ is the star's parallax; and $f_{\varpi}$ is the along-scan parallax factor. We measured the time, $t$, relative to the \gaia reference time ($t =
t_{\rm obs} - t_{\rm ref}$, with $t_{\rm ref} = 2457936.875$ for DR4).

In the \gaiamodel, the default set of free parameters is the five-parameter set \((\Delta\alpha^{\star},\Delta\delta,\mu_{\alpha^{\star}},\mu_{\delta},\varpi)\). 
Alternatively, seven- or nine-parameter models can be used to model an acceleration. The additional terms for the accelerations are
\begin{equation}
    w_{\rm acc} = 
        \left(\frac{1}{2} \alpha^{\star}{''} t^2 + \frac{1}{6} \alpha^{\star}{'''} t^3 \right) \sin{\psi} + 
        \left(\frac{1}{2} \delta'' t^2 + \frac{1}{6} \delta''' t^3 \right) \cos{\psi},
\end{equation}
corresponding to the next terms in the Taylor expansion of \(\alpha^{\star}(t)\)
and \(\delta(t)\) around zero. The seven-parameter model includes $\alpha^{\star}
{''}$ and $\delta ''$, the accelerations in equatorial coordinates. The
nine-parameter model also includes the jerks $\alpha^{\star} {'''}$ and
\(\delta'''\). This can be set in \kima through the function {\tt
set\_baseline\_model}, which takes the number of parameters in the
acceleration models as input. \edit{If an acceleration is present, it can instead be modelled as a Keplerian when constraints on the orbital parameters are desired.}

We also include a parametrised model in \kima to fit potential scan-angle-dependent signals \citep[as
discussed in][]{holl_gaia_2023-1}. This model can include the
third, fifth, and seventh scan-angle harmonics (since the first harmonic is
degenerate with parallax), with each harmonic having an amplitude \(A\) and a phase
\(\theta\):
\begin{equation}
    w_{\rm sc} = \sum_{n=3,5,7}A_{n}\cos{[n(\psi - \theta_n)]}.
\end{equation}
These terms can be included in \kima through the {\tt \edit{set\_scan\_dep\_signal}} parameter, which
can take values from zero to three. The parameter controls whether the third harmonic, the third, and fifth harmonics, or
the third, fifth, and seventh harmonics are included.

To account for planetary signals, the model can also include Keplerian components, with the number of planets, \np, a free parameter. For \gaia data, the Keplerian model is
\begin{equation}\label{eq:Gaia_kep}
    w_{\rm k} = (BX +GY)\sin{\psi} + (AX + FY)\cos{\psi},
\end{equation}
where the Thiele-Innes parameters \(A,B,F,G\) \citep{thiele_neue_1883,
van_den_bos_orbital_1926, perryman_astrometry_2018} are defined as
\begin{align}
    A &= a(\cos{\omega}\cos{\Omega} - \sin{\omega}\sin{\Omega}\cos{i}),\\
    B &= a(\cos{\omega}\sin{\Omega} + \sin{\omega}\cos{\Omega}\cos{i}),\\
    F &= -a(\sin{\omega}\cos{\Omega} + \cos{\omega}\sin{\Omega}\cos{i}),\\
    G &= -a(\sin{\omega}\sin{\Omega} - \cos{\omega}\cos{\Omega}\cos{i}).
\end{align}
Here, \(a\) is the semi-major axis of the photocentre's orbit around the centre of mass (in angular units and sometimes written as \(a_0\)), \(\omega\) is the argument of pericentre, \(\Omega\) the longitude of ascending node, and \(i\) the inclination of the orbit (where \(i=0\) corresponds to a face-on orbit). The parameters X and Y in \eq{eq:Gaia_kep} represent the orbital position in elliptical rectangular coordinates:
\begin{align}
    X &= \cos{E} - E,\\
    Y &= \sqrt{1-e^2}\sin{E},
\end{align}
where \(E\) is the eccentric anomaly and \(e\) is the eccentricity of the orbit. \edit{The free parameter that sets the orbital phase in \kima is the mean anomaly at the reference time \(\mathcal{M}_0\).}

The \kima implementation of this Keplerian model includes the option to sample either the Thiele-Innes parameters or the orbital (sometimes called geometric) parameters through the setting {\tt thiele\_innes}, which defaults to {\tt False}. When using the geometric parameters, we parametrise the inclination as \(\cos{i}\), since a uniform prior in \(\cos{i}\) corresponds to an isotropic distribution\edit{ Using Gaussian or uniform priors on the Thiele-Innes parameters leads to an effective inclination prior that favours face-on orientations. We do not recommend using the Thiele-Innes parameters for model comparison (see \sect{sec:TI_comp}), but we include this option for completeness}. The number of Keplerians, \np, can be a free parameter, with all Keplerians sharing the same prior distributions. As in the default radial velocity \kima model \citep{faria_kima_2018},  the {\tt known-object} mode also allows Keplerians with individually defined priors \edit{for a specified number of \texttt{known-objects} \(N_{\rm KO}\)}. For these components, sampling is always performed on the geometric parameters rather than on the Thiele-Innes parameters.

The full model is therefore
\begin{equation}
    w_{\rm model} = w_{\rm ss} + w_{\rm acc} + w_{\rm sc} + \sum_{i=0}^{\np} w_{{\rm k},i} +\sum_{j=0}^{N_{\rm KO}}w_{{\rm k},j},
\end{equation}
with each term, apart from $w_{\rm ss}$, calculated only when activated in the model setup.
The model is then compared to the data \edit{using} a likelihood \edit{function}, using either a Gaussian distribution,
\begin{equation}
    \mathcal{L} = \frac{1}{\sqrt{2\pi\sigma^2}} 
                  \exp\left[-\frac{(w_{\rm obs}-w_{\rm model})^2}{2\sigma^2} \right],
\end{equation}
or a Student's {\it t} distribution,
\begin{equation}
    \mathcal{L} = \frac{\Gamma\left(\frac{\nu+1}{2}\right)}%
                       {\sqrt{\pi\nu\sigma^2} \, \Gamma \left( \frac{\nu}{2} \right) }
                  \left[ 1+\frac{(w_{\rm obs} - w_{\rm model})^2}{\nu\sigma^2}\right]^{-\frac{\nu + 1}{2}}.
\end{equation}
For each datum, the variance is \(\sigma^2 = \sigma_{\rm obs}^2 + J^2\), where \(\sigma_{\rm obs}\) is the observational uncertainty on that datum, \(J\) is the astrometric jitter, and \(\nu\) is the shape parameter of the Student's {\it t} distribution. The Student's t likelihood can be used when the data contain potential outliers or the residual noise has heavy tails \citep[e.g.][]{brewer_fast_2015,agol_refining_2021,baycroft_bebop_2025}. In the absence of significant outliers, $\nu$ is large and the likelihood approximates a Gaussian distribution.

 \tbl{tab:default_priors} lists the default priors for all model parameters; users can access and modify each prior.

\subsection{\rvgaiamodel}

The \rvgaiamodel inherits all the functionality of the \gaiamodel described above and combines it with constraints from radial velocities, for which the Keplerian model is \citep[e.g.][]{hilditch_introduction_2001,lovis_radial_2010}
\begin{equation}\label{eq:RV_kep}
    V_{r} = V_{\rm sys} + K(\cos{(\omega + f)} + e\cos{\omega}),
\end{equation}
where \(\omega\) is the argument of pericentre, \(f\) is the true anomaly, and \(V_{\rm sys}\) is the systemic velocity.

Including the radial velocity data does not add any free parameters to the Keplerian: $e$ and $\omega$ are shared between the two, while \(f\) from \eq{eq:RV_kep} and \(X\) and \(Y\) from \eq{eq:Gaia_kep} depend on the shared parameters \(\mathcal{M}_0\) (the mean anomaly at the reference time) and \(P\) (the orbital period). The radial velocity semi-amplitude \(K\) and the photocentre semi-major axis \(a\) are directly related through
\begin{equation}
    K = \frac{a}{\varpi}\frac{2\pi\sin{i}}{P}\sqrt{1-e^2}.
\end{equation}
Although the reverse approach is possible, we chose to sample in \(a\) and convert to \(K\), since \(K\) has a stronger degeneracy with inclination. Because of this, the \rvgaiamodel does not use the Thiele-Innes parametrisation.

The \rvgaiamodel allows multiple radial velocity time series to be included (such as measurements from different instruments) with offsets between the datasets as free parameters and separate jitters for each instrument. It also allows a polynomial trend of up to cubic order to be fitted: 
\begin{equation}
    V_{r} = S(t-t_{\rm mid}) + Q(t-t_{\rm mid})^2 + C(t-t_{\rm mid})^3,
\end{equation}
where \(S\), \(Q\), and \(C\) are the slope, quadratic, and cubic coefficients (the free parameters being fitted), \edit{and \(t_{\rm mid}\) is the mid-point of the radial velocity baseline}.

The only shared parameters between the radial velocity and astrometry are the Keplerian orbital parameters. The jitter terms, Student's {\it t} degrees of freedom, and astrometric accelerations, and radial velocity trend parameters are not shared.

\tbl{tab:default_priors} shows the default priors for the \rvgaiamodel.
The only meaningful difference from the \gaiamodel is the prior for
$\Omega$, which now spans the range $[0,2\pi]$, since the radial velocities break the
\edit{symmetry} around $\pi$ \citep{ranalli_astrometry_2018}.

\subsection{\texttt{RVHGPMmodel}}

Here we briefly describe a model previously implemented in \kima \edit{that combines} radial-velocity data with constraints from the \textit{Hipparcos}-\gaia proper-motion anomaly \citep{brandt_hipparcos-gaia_2021,kervella_stellar_2022}. This uses the Hipparcos-Gaia catalogue of accelerations \citep{brandt_hipparcos-gaia_2021} and directly implements the method presented in \citet{venner_true_2021}. These analyses can constrain the orbital inclination and therefore the true masses of the companions. A more detailed description will be provided in \citetext{Ceva et al. in prep.}.

\section{Testing the models}\label{sec:testing}

This section details the performance of the astrometric models using real and simulated datasets. To input \gaia data into \kima through the {\tt GAIAdata} class, the data must be provided in a file with the following five columns, in order: 
\begin{itemize}
    \item time, \(t_{\rm data}\) (in tcb\_JD)
    \item along-scan position, \(w_{\rm obs}\) (in mas)
    \item along-scan position uncertainty, \(\sigma_{\rm obs}\) (in mas)
    \item scan angle, \(\psi\) (in rad)
    \item along-scan parallax-factor, \(f_{\varpi}\)
\end{itemize}
\edit{The time unit, tcb\_JD, is expressed in barycentric coordinate time in Julian days}. \edit{All this information will be provided per target in} \gaia DR4, \edit{although some processing may be necessary, such as converting times from nanoseconds to days and scan angles from degrees to radians}\footnote{\edit{The DR4 pre-release information is available at \url{https://www.cosmos.esa.int/web/gaia/dr4-prerelease}}}.

In the following, the data were binned by epoch, averaging the expected nine individual CCD measurements into a single inverse-variance-weighted  measurement, \edit{and performing outlier rejection}\footnote{\edit{See \url{https://dace.unige.ch/openData/?record=10.82180/dace-u2oj7sbb}, more details will be available in the \gaia DR4 papers}}. The \gaiamodel and \rvgaiamodel can also analyse the per-CCD data, but this requires more iterations of the numerical method to solve Kepler's equation for each proposed sample and consequently increases computational cost.

The data analysed comprise simulated data\footnote{available at \url{https://dace.unige.ch/openData/?record=10.82180/dace-gaia-ohp}} and real data for \gaia BH3 \citep{gaia_collaboration_discovery_2024}. The simulated data were generated using residuals from real \gaia data processed with the outlier-rejection and per-transit-binning procedures used by the official non-single-star pipeline \citep{halbwachs_gaia_2023}. New signals were injected into these residuals, including five-parameter astrometry and, where applicable, Keplerian signals. The data therefore have realistic cadence and noise properties of \gaia DR4 data\footnote{\edit{See \url{https://dace.unige.ch/openData/?record=10.82180/dace-gaia-ohp}.}}.

In \edit{some of} the following tests, we used non-default priors for selected parameters to improve sampling efficiency. For example, we selected the parallax prior is chosen based on the true or injected value, \edit{and we used weakly informative priors for the offsets and proper-motions of \gaia BH3. Additionally,} the eccentricity prior follows an approximation \citep{kumaraswamy_generalized_1980} to a \(\beta\) distribution, as favoured by \citet{kipping_parametrizing_2013} based on the radial-velocity planet population.

\subsection{Simulated planets}

We first applied the \gaiamodel to three simulated planet signals (Targets 1, 5, and 6 in the {\tt dace-gaia-ohp} dataset). The first is a long-period signal (\(P\approx 2000\) days), with a period close to the data time span and a high signal-to-noise ratio (S/N). The other two cases are shorter period signals (\(P\approx 90\) days), with high and low S/N, respectively. We additionally \edit{used} a simulated dataset with no planet (Target 2 in the {\tt dace-gaia-ohp} dataset) for testing compatibility limits (\sect{sec:compat})\edit{. We also injected a two-Keplerian signal into this dataset and generated radial velocities, which we jointly analysed with the \rvgaiamodel}.

\subsubsection{High-S/N long-period signal}

\begin{figure}
    \centering
    \includegraphics[width=0.9\columnwidth]{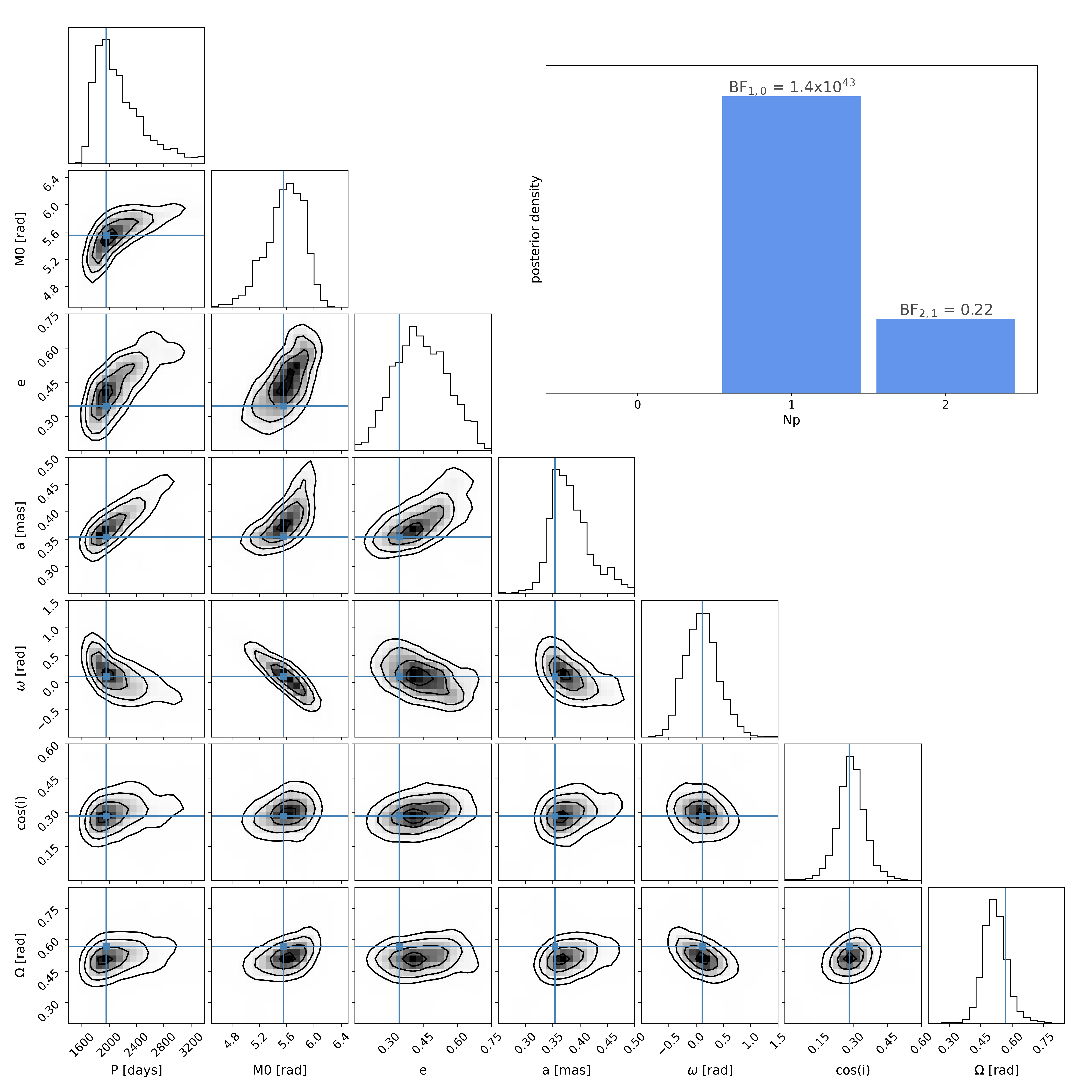}
    \caption{Corner plot of the orbital parameters for Target 1, with the injected values shown. \edit{Top right: Histogram of the posterior density for each number of Keplerians and the Bayes factors calculated by comparing each model with the model containing one fewer Keplerians.}}
    \label{fig:Target1_corner}
\end{figure}

We analysed Target 1 (the high-S/N and long-period planetary signal) with \kima, allowing the number of planets to range up to two. \edit{The S/N of the injected signal is 86.5 \citep[calculated as in][]{baycroft_predicting_2026}.} The results favour a \(\np = 1\) model over a \(\np = 0\) model with a Bayes factor of \(BF_{1,0} = 1.4\times10^{43}\). There is no evidence for a \(\np=2\) model over the \(\np=1\) model, with a Bayes factor \(BF_{2,1} <1\). \fig{fig:Target1_corner} shows these values alongside a histogram of the number of posterior samples for each number of planets, showcasing how treating the number of Keplerians as a free parameter enables model comparison.

\begin{figure}
    \centering
    \includegraphics[width=0.9\columnwidth]{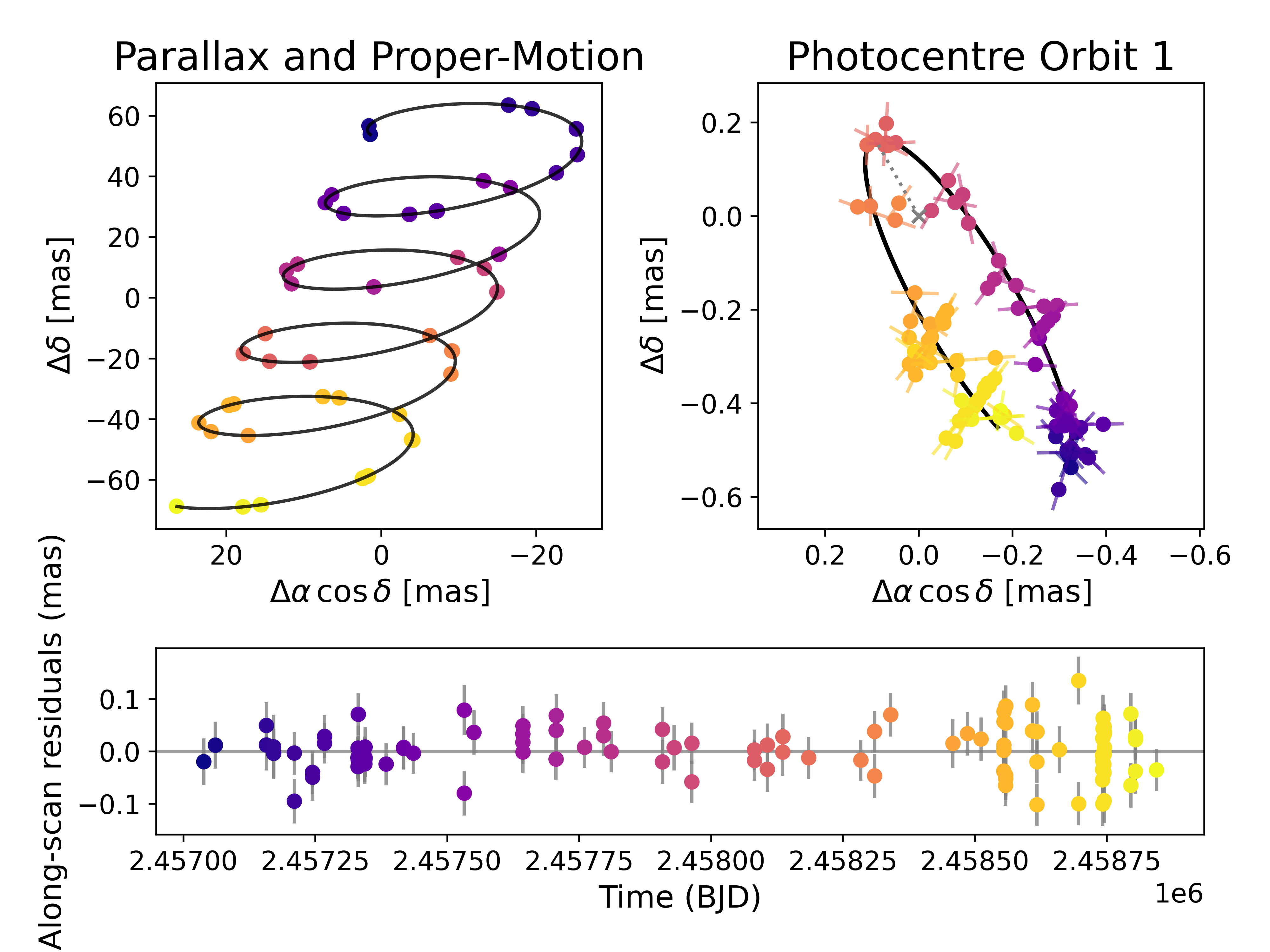}
    \caption{Phase plot of the maximum-likelihood model for Target 1. Top-left: Parallax and proper-motion with the orbital signal removed. Top-right: Orbit of the photocentre around the centre of mass. Bottom: Residuals in the along-scan direction. The colour corresponds to the time.}
    \label{fig:Target1_phase}
\end{figure}

We show the maximum-likelihood solution in \fig{fig:Target1_phase}, which was generated with \kima's post-processing {\tt phase\_plot} function\footnote{We approximated the RA and Dec values, which were not provided with the simulated data, so that the parallax and proper-motion plot can be generated. \edit{We do this by comparing the \gaia scanning law to the along-scan parallax factors.}}. The parallax and proper-motion panel uses {\tt pystrometry} \citep{johannes_sahlmann_2019_3515526}.
\fig{fig:Target1_corner} shows the posterior distributions for the planet parameters from all samples with \(\np=1\), together with the simulated values. All posterior distributions are consistent with the true values. Because the simulated orbital period is almost exactly equal to the data time span, the posterior has a long tail towards larger periods. The maximum-likelihood solution has a longer period than the injected value, \edit{so the orbit in \fig{fig:Target1_phase} is not closed}; however, \edit{it is within the \(68\%\) confidence interval and}the mode of the distribution is at the injected value.

\subsubsection{High-S/N short-period signal}

We analysed Target 5, the high-S/N, short-period planetary signal, with \kima, again allowing the number of planets to range up to two. \edit{The S/N of the injected signal is 40.4 \citep[calculated as in][]{baycroft_predicting_2026}.} The results favour a \(\np = 1\) model over a \(\np = 0\) model, with a Bayes factor of \(BF_{1,0} = 1.6\times10^{19}\). There is no evidence for a \(\np=2\) model over the \(\np=1\) model (Bayes factor \(BF_{2,1} <1\)).

We show the maximum-likelihood solution in \fig{fig:Target5_phase}. \fig{fig:Target5_corner} shows the posterior distributions of the planetary parameters from the \(\np=1\) model, together with the injected values. The values of \(\omega\) and \(\Omega\) appear incorrect. This is a consequence of limiting the range of \(\Omega \in [0,\pi]\). Astrometry alone is \edit{symmetric} under rotations of \(\Omega\) by \(\pi\) since motion out of the sky-plane is not observed, so allowing for the full range would lead to multimodal posteriors in \(\Omega\) and \(\omega\), which is defined relative to \(\Omega\). Therefore, to simplify sampling and interpretation, the default prior\footnote{The user may modify the priors as required.} in \kima limits the range of \(\Omega\), recognising that both values may differ by \(\pi\), as in this case. \fig{fig:Target5_corner2} shows the same posteriors but with \(\Omega \rightarrow \Omega + \pi\) and \(\omega \rightarrow \omega - \pi\). The solution is then consistent with the injected values.

\subsubsection{Low-S/N short-period signal}\label{sec:lowsnr}

We analysed Target 6 (the low-S/N, short-period planetary signal) with \kima, allowing the number of planets to range up to two. \edit{The S/N of the injected signal is 7.6 \citep[calculated as in][]{baycroft_predicting_2026}.} The results favour a \(\np = 1\) model over a \(\np = 0\) model, with a Bayes factor of \(BF_{1,0} \approx 100\), which lies on the boundary between strong and conclusive evidence. There is no evidence for an \(\np=2\) model over the \(\np=1\) model (Bayes factor \(BF_{2,1} <1\)).

We show the maximum-likelihood model in \fig{fig:Target6_phase}.  \fig{fig:Target6_corner} shows the posterior distributions of the planet parameters for the \(\np=1\) model, together with the injected values. Although the injected values are within the posterior distributions, the analysis favours a region with a greater inclination and a lower \(\Omega\). The values of \(P\) and \(a\) are also underestimated and overestimated, respectively. Restricting the range to \(\pi\leq\Omega\leq4\), we obtain the maximum-likelihood model within this range shown in \fig{fig:Target6_phase_correct}. We verified that this result is not a bias of \kima by analysing the data with \kepmodel \citep{delisle_analytical_2022}, which gives results consistent with those of \kima. Degeneracies such as this remain poorly explored in \gaia data. Therefore, care will be needed in the low-S/N regime, and tools such as \kima that thoroughly explore the parameter space will be important.

\subsection{Comparing the geometric and Thiele-Innes parametrisations}\label{sec:TI_comp}

We tested the Thiele-Innes parametrisation. We fitted Target 5 using the Thiele-Innes parameters, fixing the number of planets to \(\np=1\).  \fig{fig:param_comp_TI} shows the posterior distributions in \(P,a,e,\omega,\cos{i},{\rm\,and\,}\Omega\), demonstrating that the model recovers consistent posterior distributions.

\begin{figure}
    \centering
    \includegraphics[width=0.9\linewidth]{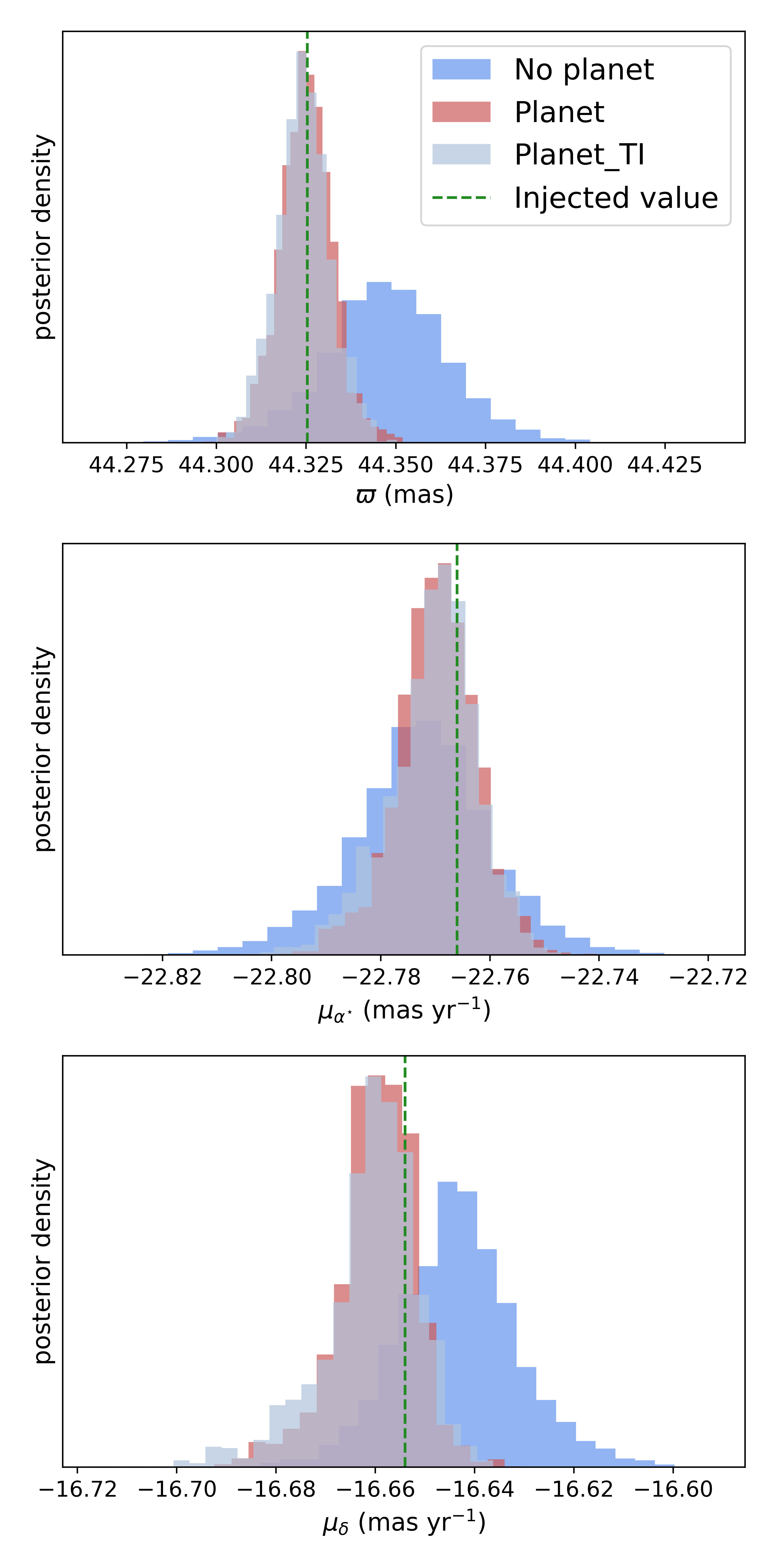}
    \caption{Comparison of the posterior distribution of parallax \(\varpi\)
             (top) and proper motions \(\mu_{\alpha^\star}\) (middle) and
             \(\mu_\delta\) (bottom) for runs with zero and one planet. The injected values are also shown. }
    \label{fig:parameter_comp}
\end{figure}

For this case, we also show how the inclusion of the planet improves the baseline astrometry constraints. \fig{fig:parameter_comp} shows the posterior distributions obtained when fitting a planet (using both the geometric and Thiele-Innes parametrisations) compared with fitting no planet. The results from both planet fits are more precise and closer to the injected values, even though this case has an orbit much shorter than the data time span and not close to one year, so strong degeneracies are not expected. In particular, the parallax precision increases by more than a factor of two.

Although the posteriors on the various parameters are consistent, the Bayes factor is smaller when using the Thiele-Innes parameters, with a value of \(BF_{1,0} = 6.5\times10^{17}\), compared to the value of \(BF_{1,0} = 1.6\times10^{19}\) for the original fit. Directly comparing the two parametrisations using a Bayes factor is not particularly meaningful for model comparison because they are effectively the same model. However, it does reveal the impact that parametrisation and the choice of priors can have on \edit{Bayesian} model comparison. This decrease is likely due to the larger prior-posterior volume ratio in the Thiele-Innes parametrisation, meaning that the `Occam factor' implicit in the Bayes factor calculation penalises it more \citep{jefferys_ockhams_1992,gregory_bayesian_2005}. \edit{This ratio is larger because each Thiele-Innes parameter has a coefficient of \(a\), which is often more tightly constrained than \(\omega,\Omega\,{\,\rm ,or\,}\cos{i}\), and can vary over orders of magnitude unlike these other parameters. Since the Thiele-Innes parameters can be positive or negative, the default prior is Gaussian, which underweighs small values of \(a\) compared with the prior in the geometric parametrisation, which is a modified log-uniform distribution. This suggests that the Bayes-factor discrepancy would be worse for a smaller signal. We confirmed this by analysing the data for the low-S/N Target 6 using the default priors. The geometric parametrisation gives a Bayes factor of \(BF_{1,0} \approx 100\) (as in \sect{sec:lowsnr}), whereas the Thiele-Innes parametrisation gives a Bayes factor of \(BF_{2,1} \approx 0.01\). Therefore, especially for low-amplitude signals, we do not recommend using the Thiele-Innes parametrisation with the current default priors.}

\subsection{Scan-angle-dependent signal}

Here, we investigate whether model comparison within \kima can distinguish a true Keplerian signal from a scan-angle-dependent signal. Target 4 from {\tt dace-gaia-ohp} is a scan-angle-dependent signal with a period close to 90 days, as one might expect from a marginally resolved binary star \citep[see][]{holl_gaia_2023-1}. Keplerian signals were injected into Targets 5 and 6 as functions of time, rather than scan angle, although this may not be distinguishable since their time dependence matches the scan angle's well. We used \kima to perform Bayesian model comparison on these to determine whether the scenarios can be distinguished.

For each of the three datasets we performed five separate \kima analyses: one with only the baseline five-parameter astrometry; one with a single Keplerian\footnote{Here we fixed the number of Keplerians to one so that we only considered the one-Keplerian model in the evidence calculation.}; one with the scan-angle-dependent signal including only the third harmonic; one including the third and fifth harmonics; and finally one including the third, fifth, and seventh harmonics. We used the same priors in all calculations and we ran\edit{each} analysis for \edit{50\,000} steps\edit{, with convergence of the nested sampler occurring before this point in every case. } We calculated the evidence for each model and used the evidence ratio (Bayes factor) for model comparison.

\begin{table}
    \renewcommand{\arraystretch}{1.1}
    \caption{Estimated $\log Z$ for different models in the analysis of three targets with signals around 90 days.}
    \label{tab:ModelCompLogZ}
    \centering
    \begin{tabular}{c|ccc}
        Model & Target 4 & Target 5 & Target 6 \\
        \hline
        \(\np = 0,\,N_h = 0\) & 42.40 & 31.56 & 107.78 \\
        \(\np = 1,\,N_h = 0\) & 67.56 & 75.90 & \underline{112.53} \\
        \(\np = 0,\,N_h = 1\) & 59.01 & 68.93 & 108.91 \\
        \(\np = 0,\,N_h = 2\) & \underline{74.02} & \underline{79.77} & 105.92 \\
        \(\np = 0,\,N_h = 3\) & 73.17 & 77.30 & 103.50 \\
    \end{tabular}
    \tablefoot{The quantity $\log Z$ is the natural logarithm of the Bayesian evidence. The parameter \np is the number of Keplerians include, while \(N_h\) denotes the number of scan-angle harmonics included. In each case, the largest value is underlined.}
    \renewcommand{\arraystretch}{1}
\end{table}

Target 4 is the injected scan-angle signal, and the preferred model is the scan-angle signal with two harmonics included (third and fifth). It has a Bayes factor of \(BF \approx 640\) relative to the \(\np=1\) model, providing decisive evidence for a scan-angle-dependent signal rather than a Keplerian \citep[the typical threshold for overwhelming evidence in favour of a model is a Bayes factor between 100 and 140;][]{kass_bayes_1995,trotta_bayes_2008}. The model without signal and the model with a single harmonic can both be rejected. The model with three harmonics is marginally disfavoured compared with the two-harmonic model. Although there is no strong evidence distinguishing the two models, the two-harmonic model is simpler. \fig{fig:Target4_phase} shows the phase plot for the maximum-likelihood solution within the two-harmonic model.

Target 5 is the high-S/N Keplerian that we recovered above. However, the two-harmonic scan-angle model again has the highest evidence, with a Bayes factor of \(BF \approx50\) relative to the \(\np=1\) model. Although this is not decisive evidence, it would be classified as strong evidence in favour of a scan-angle-dependent signal rather than a true orbit. \fig{fig:Target5_phase_sa} shows the phase plot for the maximum-likelihood solution within the two-harmonic model. Despite being injected as a Keplerian, this large-amplitude signal can be well reproduced with a two-harmonic scan-angle-dependent signal. \edit{The two-harmonic model has three fewer free parameters than the Keplerian model. Therefore, when a Keplerian is not distinguishable from the functional form of harmonics, the Bayes factor is expected to favour the simpler model.}

Target 6 is the low-S/N Keplerian, and despite this, the analysis favours a Keplerian origin for the signal with a Bayes factor of \(BF \approx 40\) relative to the best scan-angle-dependent model (the one-harmonic model). This is strong evidence for a planetary origin of the orbit. The one-harmonic model is only marginally better than the five-parameter astrometric model and would therefore not be favoured over a no-signal model. \fig{fig:Target6_phase_sa} shows the phase plot for the maximum-likelihood solution within the one-harmonic model. This signal is not very well modelled by the scan-angle-dependent model, which is consistent with a straight line. \edit{Using the Bayesian information criterion to compare the models, based on analysis with \kepmodel, favours the same model in each case as the \kima Bayes factor comparison.}

Therefore, \kima has the ability to model potential sources of false-positive signals in \gaia data originating from scan-angle-dependent signals arising from marginally resolved binaries. In some cases it can be used to distinguish true planets from false positives, but not in all cases. Follow-up observations may therefore be necessary, such as high-resolution \edit{spectroscopy or} imaging, to assess the presence of such a binary.

\subsection{\edit{Multiple signals}}

\edit{To further demonstrate its functionality, we used \kima on a simulated two-Keplerian dataset based on the Target 2 time series, using \gaia data and radial velocities. We simulated one signal with a high astrometric S/N of 110 that is exactly face-on and therefore should have no influence on the radial velocities, and a second signal with a lower S/N of 13.4. We simulated 40 radial-velocity measurements spanning approximately two years, with a precision of around 6 m/s. The low-astrometric-S/N signal has a radial-velocity semi-amplitude \(K=7.1\) m/s.}

\edit{We performed the analysis using the known-object mode for the high-S/N signal to improve sampling efficiency, together with up to three additional Keplerians. Using the \gaiamodel, the Bayes factors are \(BF_{2,1} = 9.8\times10^6\), \(BF_{3,2} = 3.3\), and \(BF_{4,3} = 0.35\); using the \rvgaiamodel, they are \(BF_{2,1} = 5.1\times10^9\), \(BF_{3,2} = 0.74\), and \(BF_{4,3} = 0.62\). Therefore, the \gaia data alone provide overwhelming evidence for the second Keplerian, which becomes stronger with the inclusion of the radial velocities.}

\edit{We show the posterior distributions for \(P\), \(a\), and \(\cos{i}\) for each of the two signals in \fig{fig:multibody}, together with the injected values and the parameters obtained from \kepmodel. The results are consistent with the injected values, and the peaks of the distributions obtained from the fitting \gaia data alone agree closely with the \kepmodel optimisation, except for the astrometric semi-major axis of the low-S/N signal, which \kepmodel overestimates. Including an uncertainty estimate for the \kepmodel solution gives \(a = 0.095  \pm  0.038\), making it marginally compatible with the \kima results. This demonstrates the advantage of a fully Bayesian analysis in low-S/N situations. Including the radial velocities using the \rvgaiamodel slightly improves most parameters, particularly the inclination. For the face-on signal (a 20 \(M_{\rm Jup}\) companion with a period of 221 day at 32 pc), the radial velocities constrain the inclination at the \(90^{\rm th}\) percentile from \(\leq19.5^{\circ}\) to \(\leq0.26^{\circ}\).}

\subsection{Gaia BH3}

Finally, we applied the \gaiamodel to real \gaia astrometric data for the \gaia BH3 system and then applied the \rvgaiamodel to the \gaia astrometric and radial-velocity data published in \citep{gaia_collaboration_discovery_2024}. We compared the results with those obtained using \kepmodel \citep{delisle_analytical_2022}, following the analysis published by the European Space Agency\footnote{See \href{https://github.com/esa/gaia-bhthree}{github.com/esa/gaia-bhthree}.}. \editt{We show the phase and corner plots for the \gaiamodel analysis, compared with the \kepmodel results, in \fig{fig:BH3_GAIA_phase} and \fig{fig:BH3_GAIA_corner}. We show the astrometric phase plot, radial-velocity phase plot, and corner plot for the \rvgaiamodel analysis in \fig{fig:BH3_RVGAIA_phase_ast}, \fig{fig:BH3_RVGAIA_phase_rv}, and \fig{fig:BH3_RVGAIA_corner}, respectively. These show that the results from \kima are consistent with the ESA \kepmodel analysis}

The \gaia BH3 signal is very large (S/N \(\gtrsim 3000\)), which means that the region of high likelihood in the parameter space is quite narrow. Therefore, during a blind search for the signal with very wide priors, the nested sampler can struggle to identify such a sharp feature. For such strong signals, we recommend using the known-object mode  with narrower priors, as we do here. \edit{T}hese very large-amplitude signals can be detected using a blind search in \kima by fixing the astrometric jitter to a large value, thus artificially widening the likelihood peak. \edit{Another option for very high-S/N signals is to perform a periodogram search and optimisation, for example using \kepmodel, and generate the known-object priors based on the results.}

\section{Compatibility limits}\label{sec:compat}

 Radial-velocity studies have used \kima to obtain compatibility limits, which provide a Bayesian equivalent to the detection limits often obtained through injection-recovery \citep{standing_bebop_2022,john_sub-m_2023,figueira_comprehensive_2025,faria_inferring_2025}. We demonstrate this for the \gaiamodel using Target 1 from the simulated datasets and the \gaia BH3 astrometric data. We calculated the compatibility limits directly from the posterior samples generated in the \kima analysis. The posterior samples with \(\np > {\np}_{\rm ,det}\) contain Keplerians beyond those detected that are compatible with the data (hence the name). We took these posterior samples and converted \(a\) to mass to generate compatibility limits in orbital period-mass space.

\begin{figure*}[!h]
\sidecaption
    \centering
    \begin{tabular}{cc}
        \includegraphics[width=6cm]{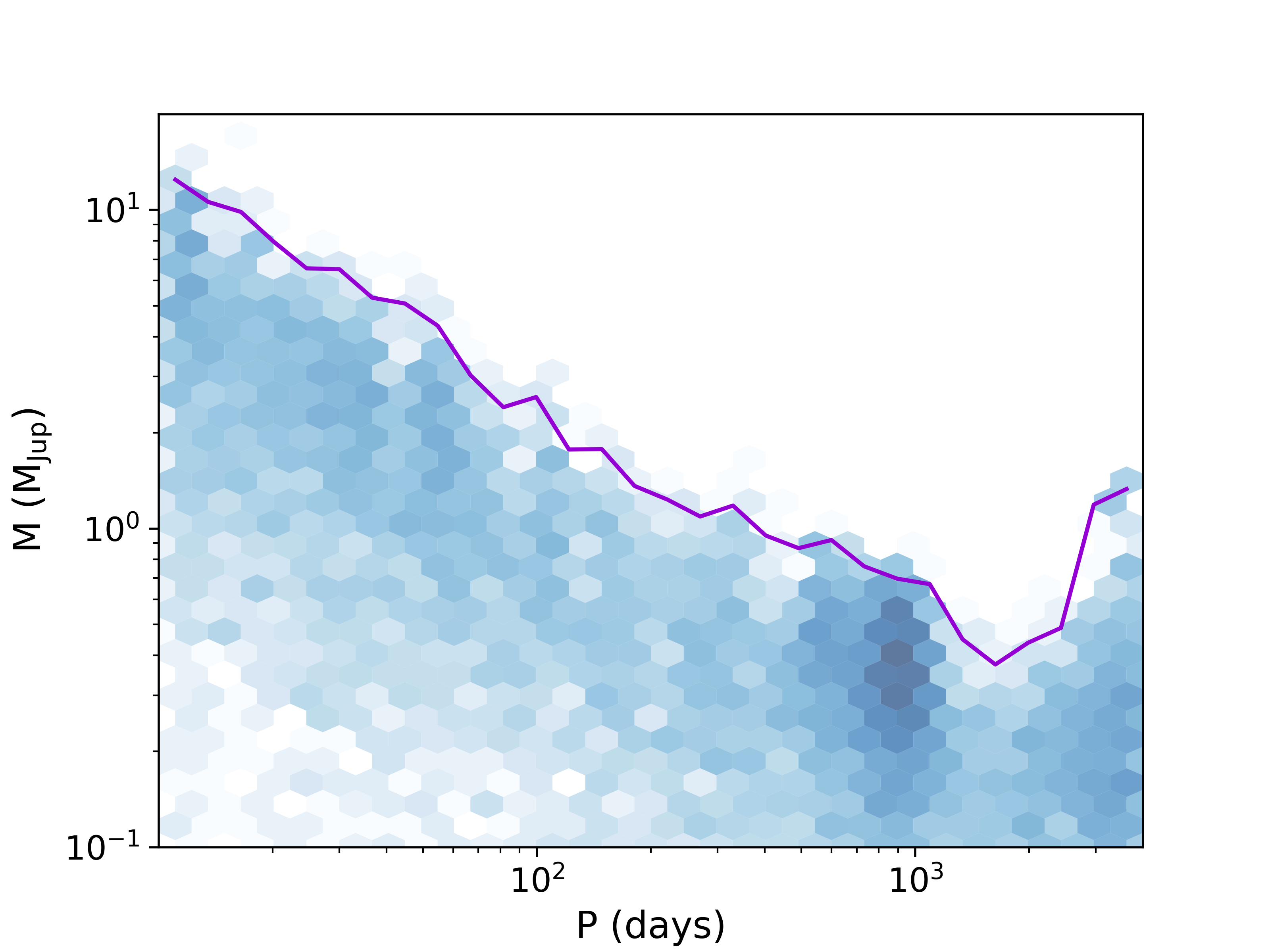} & \includegraphics[width=6cm]{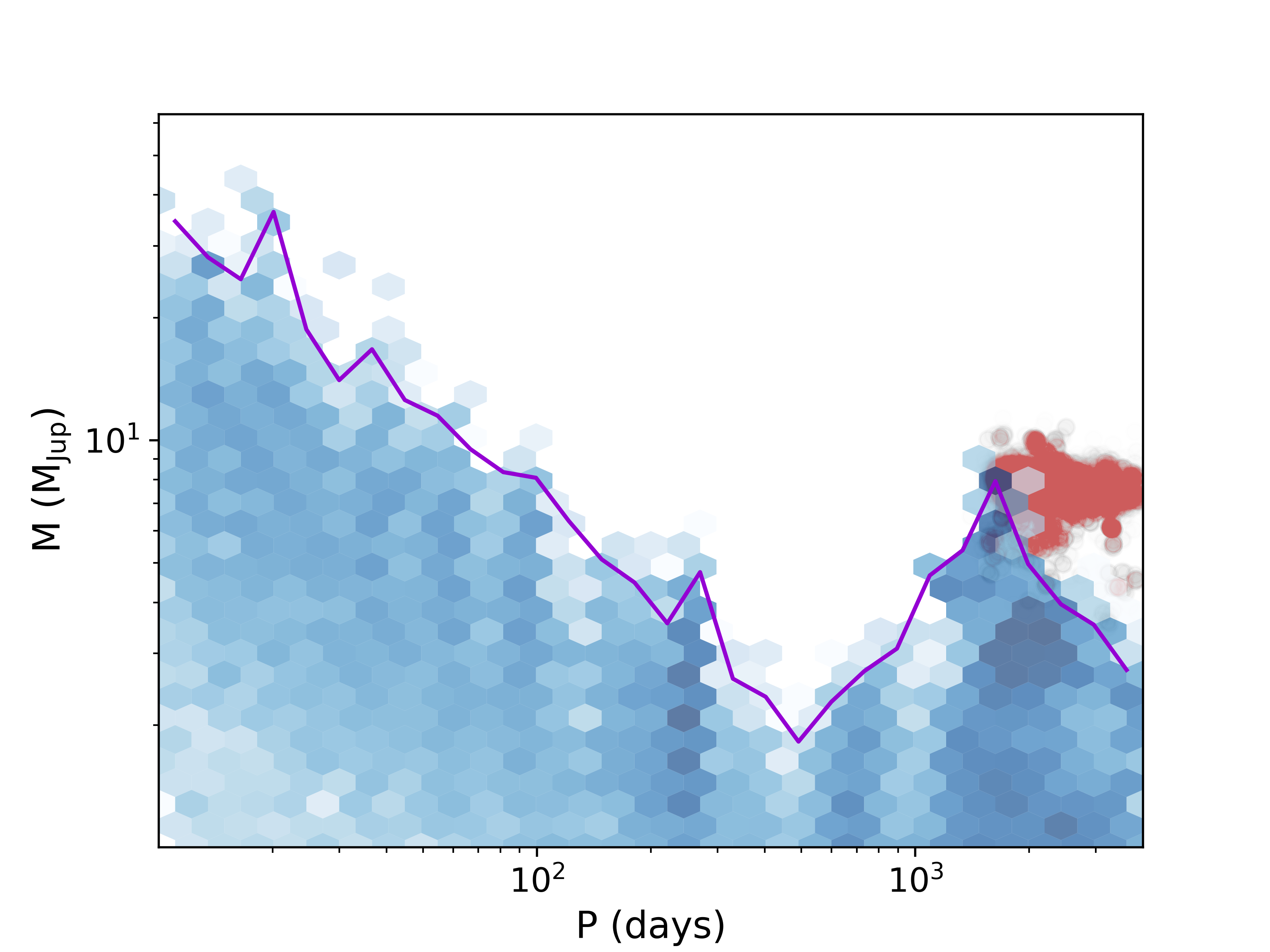} \\
        \includegraphics[width=6cm]{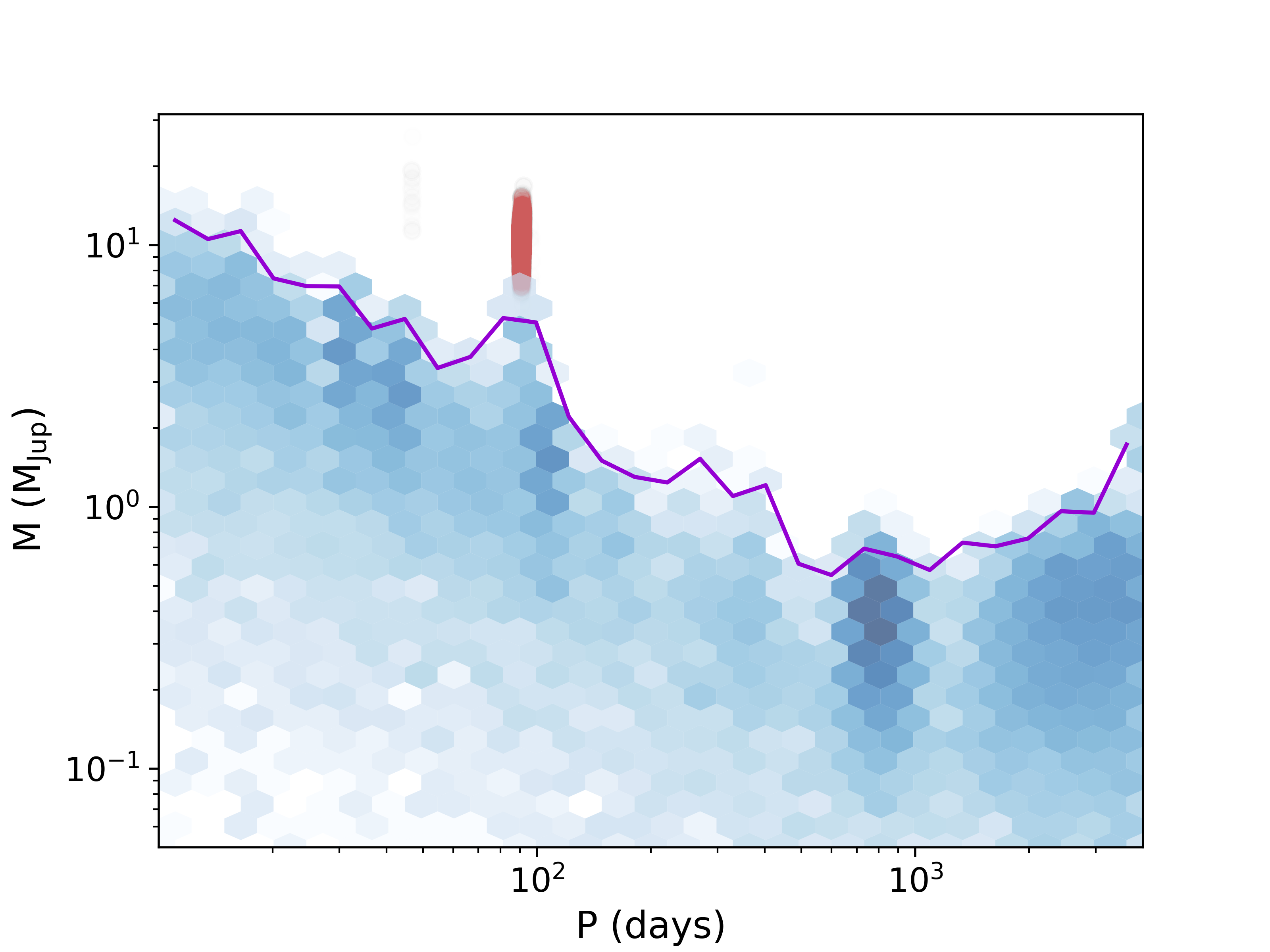} & \includegraphics[width=6cm]{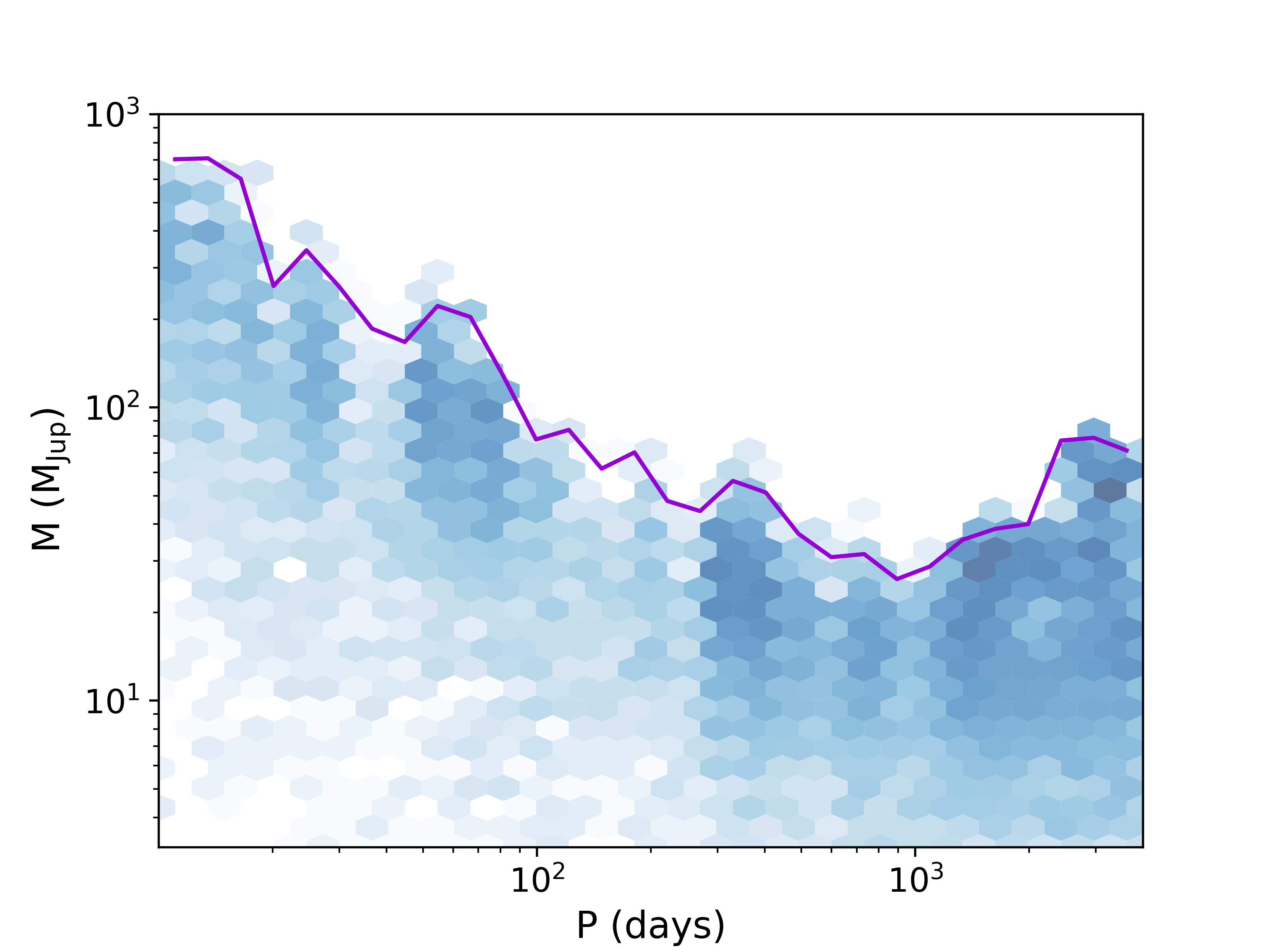} 
    \end{tabular}
    \caption{Top left: Posterior density for a second Keplerian for Target 2. Top right: Posterior density for a second Keplerian for Target 1. Posterior samples for the detected planet are shown in red. Bottom left: Posterior density for a second Keplerian for Target 5. Bottom right: Posterior density for a second Keplerian for \gaia BH3. The \(99^{\rm th}\) percentile line is shown in all panels.}
    \label{fig:compat}
\end{figure*}

\fig{fig:compat} shows the posterior density and a \(99^{\rm th}\)-percentile in each bin for Targets\,1 (top right), 2 (top left), 5 (bottom left), and \gaia BH3 (bottom right). For Targets\,1 and 5, we also show the posterior samples for the detected planets; for \gaia BH3, we do not show the orbital posteriors because the mass is very high in comparison.

The limits for Target\,2, which has no planet, show the expected shape, with a minimum around the data time span (\(\approx 2000\) days for \gaia DR4). The other three targets exhibit the same basic shape, with some deviations. For Target\,5, there is a small spike at the same orbital period as the planet. This is an expected feature, where two Keplerians of the same period can combine (analogously to interfering waves) and produce another Keplerian. For \gaia BH3, whose orbit is much longer, the orbital period is not sampled in the\kima blind search. There is a slight increase in the compatibility limit from \(\approx 1000\) days, reflecting compatibility with long-period, high-mass signals that should in principle be detectable in \gaia if they existed (i.e. a \gaia S/N above 20). This may arise because the orbit of \gaia BH3 has not been fully covered, and so by altering its parameters (especially \(\omega,\Omega,\cos{i},{\, \rm and\,}\mathcal{M}_0\)), an additional Keplerian of large amplitude can b\edit{e} fitted. Although this degeneracy implies that compatibility limits including partially covered orbits can contain large-amplitude signals, it also means that a genuine additional signal might be absorbed by the long-period signal. This \edit{may manifest similarly to the effect noted in radial velocity data by \citet{wittenmyer_anglo-australian_2012},  where a single eccentric planet was initially detected, but a second planet was subsequently detected, and its inclusion impacted the measured parameters of the first planet, notably decreasing the measured eccentricity}. 

Target\,1 shows a very large posterior spike at \(\approx 2000\) days, where the trough is expected. This is also the orbital period of the injected planet, and the two sets of posteriors overlap. This is likely a combination of the two effects described above: the injected planet's orbital period lies within the sampled range, and some of the planet's posteriors are for not fully covered orbits. As a quick test of whether both effects contribute, we performed the same analysis with an upper period bound at 2000 days. The rise that peaks between 1000 to 2000 days is still present, but less pronounced. \fig{fig:2plsamps} shows a set of two-Keplerian posteriors for Target\,1, compared to the data and the maximum-likelihood two-Keplerian model. Small-scale variations should not be over-interpreted because the posteriors have different values of the baseline astrometry that has been removed. However, this result shows that some posteriors have very different parameters for the first planet, with these differences offset by the second signal, which sometimes has a comparable amplitude.

An in-depth analysis of degeneracies in multi-object \gaia analyses is beyond the scope of this work. \citet{lammers_exoplanet_2026} present an analysis of the accuracy of solutions at different orbital periods \edit{and S/N}, while \citet{yahalomi_astrometric_2026} investigate the 2:1 period ratio degeneracy \citep[analogous to that for radial velocities;][]{anglada-escude_how_2010}.

\section{Conclusions}\label{sec:conclusions}

In this work, we provide and test the \gaiamodel and \rvgaiamodel in \kima to analyse the epoch
astrometry of \gaia and combine it with radial velocities. The models can
include a free number of Keplerians, specified (known-object) Keplerians,
astrometric acceleration terms, and scan-angle-dependent signals. We tested the
models on real and simulated \gaia data and obtain results consistent with other
established tools. We show that in some cases \kima can distinguish a
Keplerian signal from a scan-angle-dependent signal. We also demonstrate how
\kima can generate compatibility limits from \gaia data.

The \kima code is in active development and is publicly available from an open-source
repository at \href{https://github.com/kima-org/kima}{github.com/kima-org/kima},
with documentation at \href{https://www.kima.science/}{www.kima.science}. 
\begin{acknowledgements}

We thank Amaury Triaud and Johannes Sahlmann for useful conversations. 
This work has made use of the following open-source software: \texttt{numpy}
\citep{harris_array_2020}; \texttt{astropy}
\citep{astropy_collaboration_astropy_2022}; \texttt{matplotlib}
\citep{hunter_matplotlib_2007}; \texttt{nanobind} \citep{nanobind}.

\end{acknowledgements}

\bibliographystyle{aa}
\bibliography{aa61563-26}


\begin{appendix}

\section{Priors}\label{sec:priors}

This appendix lists the default priors used in the \gaiamodel and \rvgaiamodel models. Some of the priors (e.g. for the baseline astrometric parameters) are admittedly arbitrary and may not be suitable in all cases. It is also likely that these defaults may change after the release of DR4 and the \gaia data is better understood. The default priors are included so that the code can run with less manual input, but for serious analyses it is best to set priors suitable to the system under study \edit{in particular for the baseline astrometry priors}. To facilitate this, in \kima's helper package {\tt pykima}, \edit{the} function \edit{\texttt{get\_gaussian\_prior\_astrometric\_baseline}} is provided to use a linear fit to calculate the baseline 5-parameter solution and return suitable priors that can then be manually included in the model. \editt{The use of this function rather than the defaults is especially relevant for nearby stars with high proper-motions and precise data.} \edit{Depending on the science case being studied different priors on the orbital parameters may also be chosen, for example shorter periods if studying close binary stars, or large amplitudes if studying black holes.}

Note that the Known-Object parameters do not have default priors, and these must be specified by the user.

The probability distributions used are: Gaussian \(\mathcal{N}\), Uniform \(\mathcal{U}\), LogUniform \(\mathcal{LU}\), and ModifiedLogUniform \(\mathcal{U}\) which is Loguniform between the two bounds and Uniform between 0 and the lower bound (allowing to use the multi-order-of-magnitude advantages of a LogUniform distribution without excluding 0). 

\begin{table}[h]
    \renewcommand{\arraystretch}{1.15}
    \caption{Default priors for the various parameters in the GAIAmodel and RVGAIAmodel.}
    \centering
    \begin{tabular}{lll}
        \hline
        \hline
        Parameter & Units & Default prior  \\
        \hline
        \multicolumn{3}{c}{{\it Shared}}\\[0.1em]
        \multicolumn{3}{l}{{\it baseline}}\\
        \(\varpi\) & \(\rm mas\)  &  \(\mathcal{LU}(1,100)\)  \\
        \(\Delta\alpha^{\star}\) & \(\rm mas\) & \(\mathcal{N}(0,1)\)  \\
        \(\Delta\delta\) & \(\rm mas\) & \(\mathcal{N}(0,1)\)  \\
        \(\mu_{\alpha^{\star}}\) & \(\rm mas/yr\)  & \(\mathcal{N}(0,100)\)  \\
        \(\mu_{\delta}\) & \(\rm mas/yr\)  & \(\mathcal{N}(0,100)\)  \\
        \(\alpha^{\star}{''}\) & \(\rm mas/yr^{2}\)  & \(\mathcal{N}(0,2)\)  \\
        \(\alpha^{\star}{'''}\) & \(\rm mas/yr^{3}\)  & \(\mathcal{N}(0,10)\)  \\
        \(\delta''\) & \(\rm mas/yr^{2}\)  & \(\mathcal{N}(0,2)\) \\
        \(\delta'''\) & \(\rm mas/yr^{3}\)  & \(\mathcal{N}(0,10)\)  \\
        \multicolumn{3}{l}{{\it scan-angle}}\\
        \(A_n\) & \(\rm mas\)  & \(\mathcal{MLU}(0.05,10)\)\\
        \(\theta_n\) & \(\rm rad\)  & \(\mathcal{U}(0,\frac{2\pi}{2n+3})\) \\
        \multicolumn{3}{l}{{\it Keplerians}}\\
        \(P\) & \(\rm days\)  & \(\mathcal{LU}(10,\edit{10}000)\)  \\
        \(\mathcal{M}_0\) & \(\rm rad\)  & \(\mathcal{U}(0,2\pi)\)  \\
        \(e\) &   & \(\mathcal{U}(0,1)\)  \\
        \(a\) & \(\rm mas\)  & \(\mathcal{MLU}(0.01,10)\) \\ 
        \(\omega\) & \(\rm rad\)  & \(\mathcal{U}(0,2\pi)\)  \\
        \(\Omega\) & \(\rm rad\)  & $\mathcal{U}(0,\pi)$ or $\mathcal{U}(0,2\pi)$ $^{\dagger}$  \\
        \(\cos{i}\) &  & \(\mathcal{U}(-1,1)\)\\
        \multicolumn{3}{l}{{\it student-t likelihood}}\\
        \(\nu\) &  & \(\mathcal{LU}(2,1000)\)\\
        \hline\\[-0.8em]
        \multicolumn{3}{c}{{\gaiamodel \textit{only}}}\\[0.1em]
        \(A,B,F,G\) & \(\rm mas\)  & \(\mathcal{N}(0,0.5)\)  \\
        \hline\\[-0.8em]
        \multicolumn{3}{c}{{\rvgaiamodel \textit{only}}}\\[0.1em]
        \(J_{\rm Gaia}\) & \(\rm mas\)  & \(\mathcal{MLU}(0.01,10)\)  \\
        \(J_{\rm RV}\) & \(\rm m/s\)  &  \(\mathcal{MLU}(1,\frac{\Delta RV_{\rm max}}{10})\) \\
        \(V_{\rm sys}\) & \(\rm m/s\)  &  \(\mathcal{U}(RV_{\rm min},RV_{\rm max})\) \\
        RV offset & \(\rm m/s\)  & \(\mathcal{U}(-\Delta RV_{\rm max}, \Delta RV_{\rm max})\) \\
        \(S\) & \(\rm m/s/day\)  &  \(\mathcal{N}(0,\frac{2\Delta RV_{\rm max}}{\Delta t})\) \\
        \(Q\) & \(\rm m/s/day^{2}\)  & \(\mathcal{N}(0,\frac{2\Delta RV_{\rm max}}{\Delta t^2})\) \\
        \(C\) & \(\rm m/s/day^{3}\)  & \(\mathcal{N}(0,\frac{2\Delta RV_{\rm max}}{\Delta t^3})\) \\
        \hline
    \end{tabular}
    \tablefoot{ \(RV_{\rm min}\) and \(RV_{\rm max}\) are the minimum and maximum values of the radial velocities. \(\Delta RV_{\rm max}\) is the half of the difference between the maximum value of the radial velocity and the minimum value.\\ 
    \(^{\dagger}\) For the GAIAmodel the prior is \(\mathcal{U}(0,\pi)\), but for the RVGAIAmodel it is instead \(\mathcal{U}(0,2\pi)\).}
    \renewcommand{\arraystretch}{1}
    \label{tab:default_priors}
\end{table}

\section{Additional figures}

\begin{figure}
    \centering
    \includegraphics[width=0.9\columnwidth]{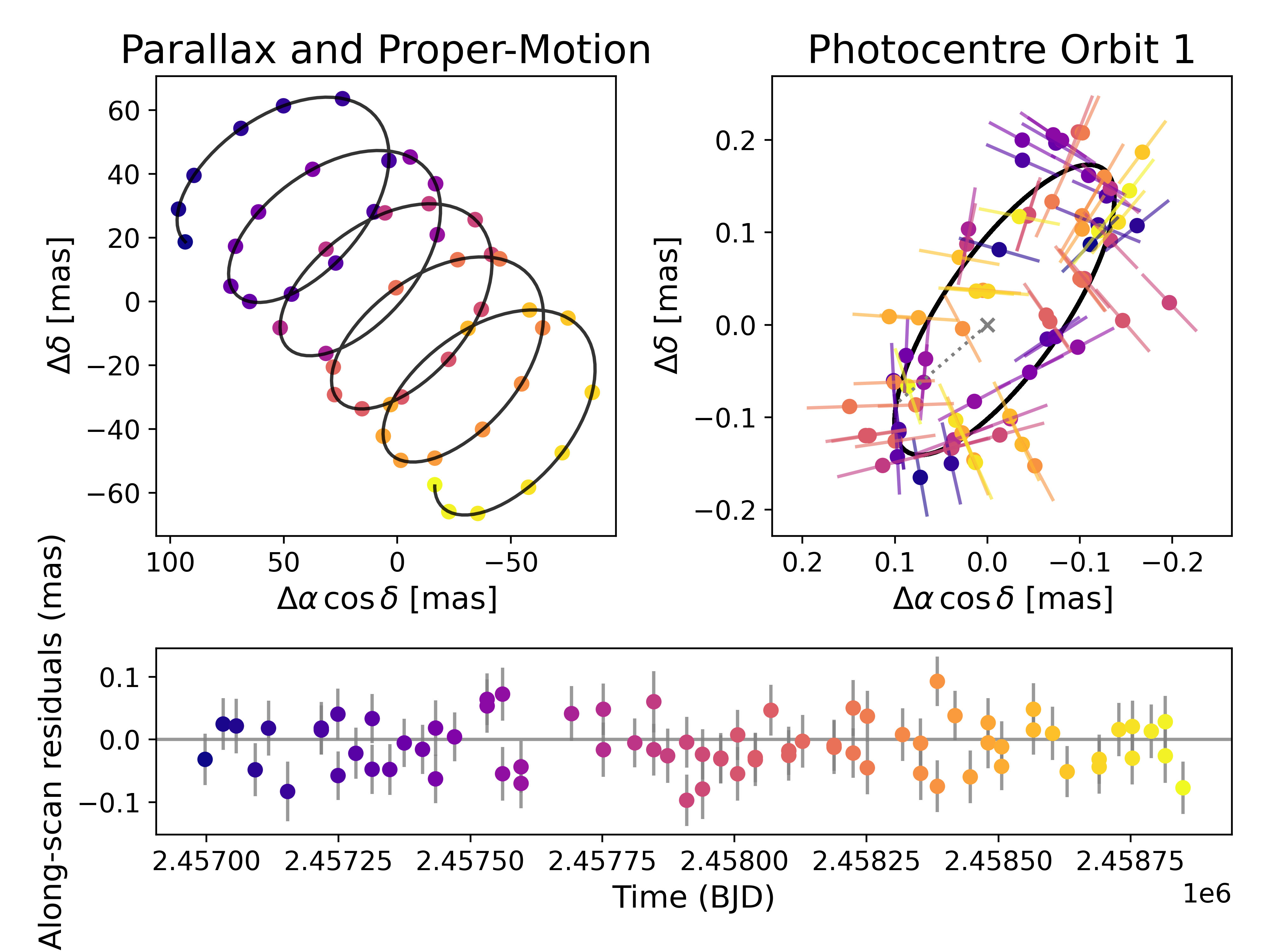}
    \caption{Phase plot of the maximum-likelihood model for Target 5. Top-left: Parallax and proper-motion with orbit signal removed. Top-right: Orbit of the photocentre around the centre-of-mass. Bottom: residuals in the along-scan direction. The colour corresponds to the time.}
    \label{fig:Target5_phase}
\end{figure}

\begin{figure}
    \centering
    \includegraphics[width=0.9\columnwidth]{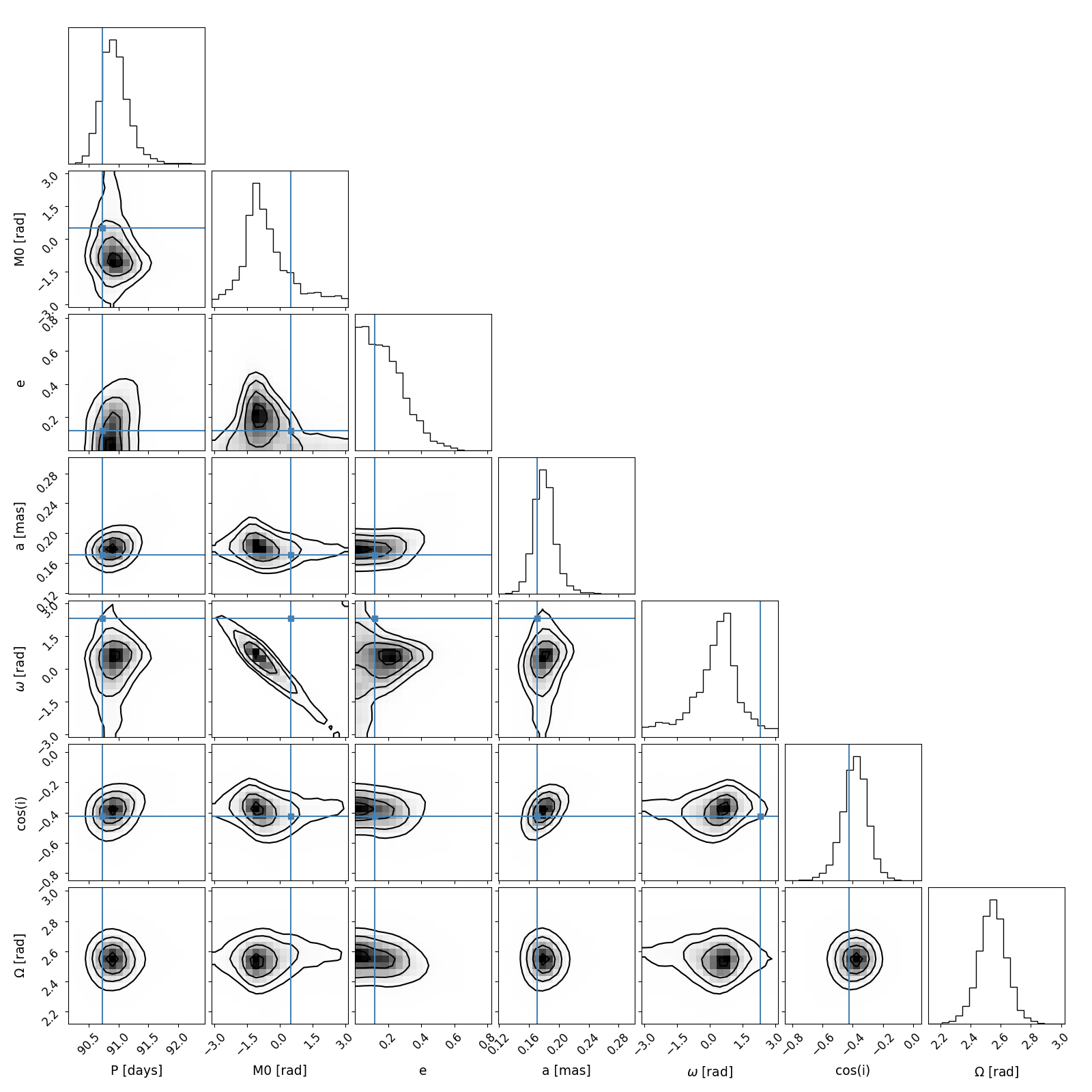}
    \caption{Corner plot of the orbital parameters for Target 5 with the injected values shown. Original values of \(\omega\) and \(\Omega\) are used.}
    \label{fig:Target5_corner}
\end{figure}

\begin{figure}
    \centering
    \includegraphics[width=0.9\columnwidth]{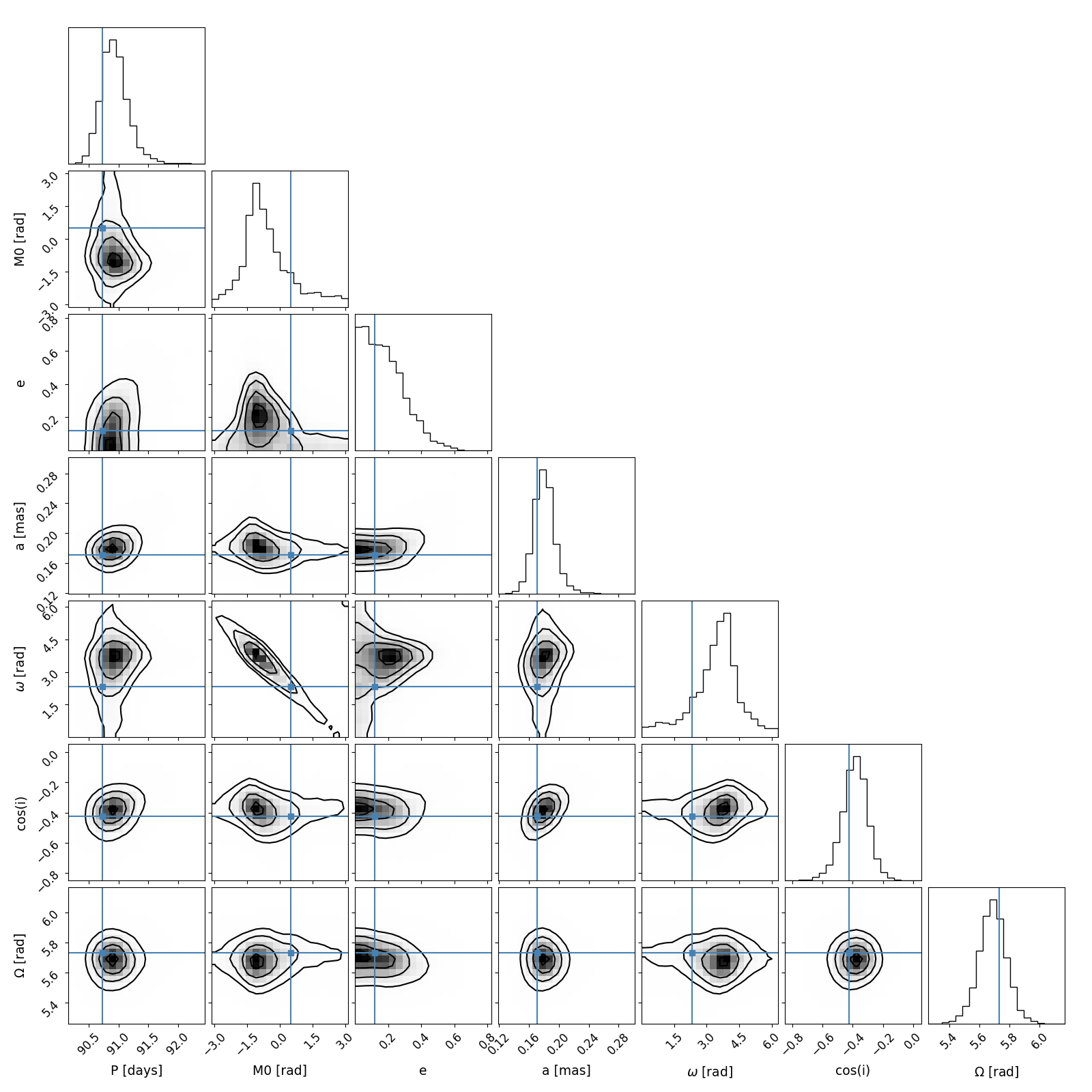}
    \caption{Corner plot of the orbital parameters for Target 5 with the injected values shown. Corrected values of \(\omega\) and \(\Omega\) are used.}
    \label{fig:Target5_corner2}
\end{figure}

\begin{figure}
    \centering
    \includegraphics[width=0.9\columnwidth]{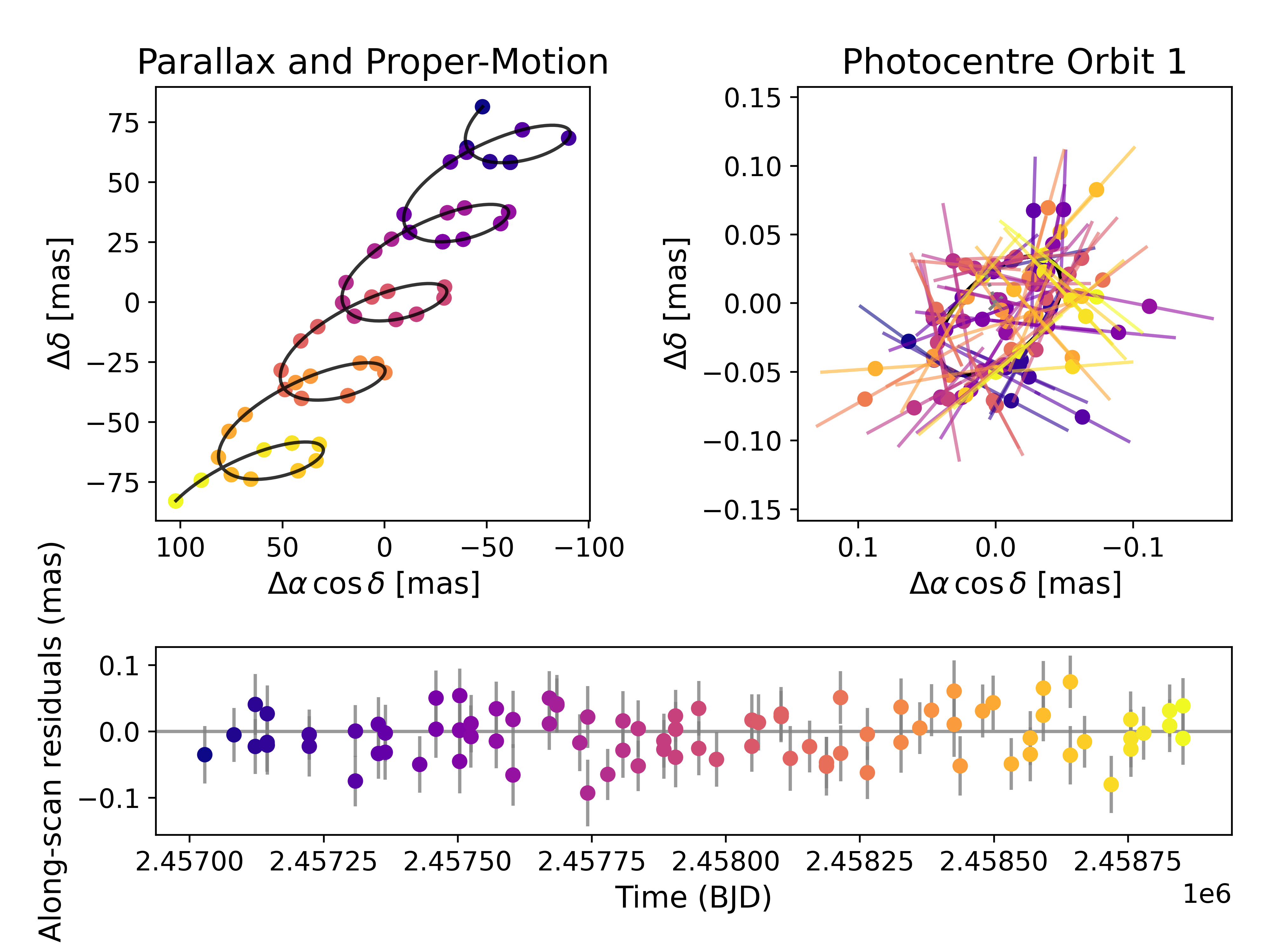}
    \caption{Phase plot of the maximum-likelihood model for Target 6. Top-left: Parallax and proper-motion with orbit signal removed. Top-right: Orbit of the photocentre around the centre-of-mass. Bottom: residuals in the along-scan direction. The colour corresponds to the time.}
    \label{fig:Target6_phase}
\end{figure}

\begin{figure}
    \centering
    \includegraphics[width=0.9\columnwidth]{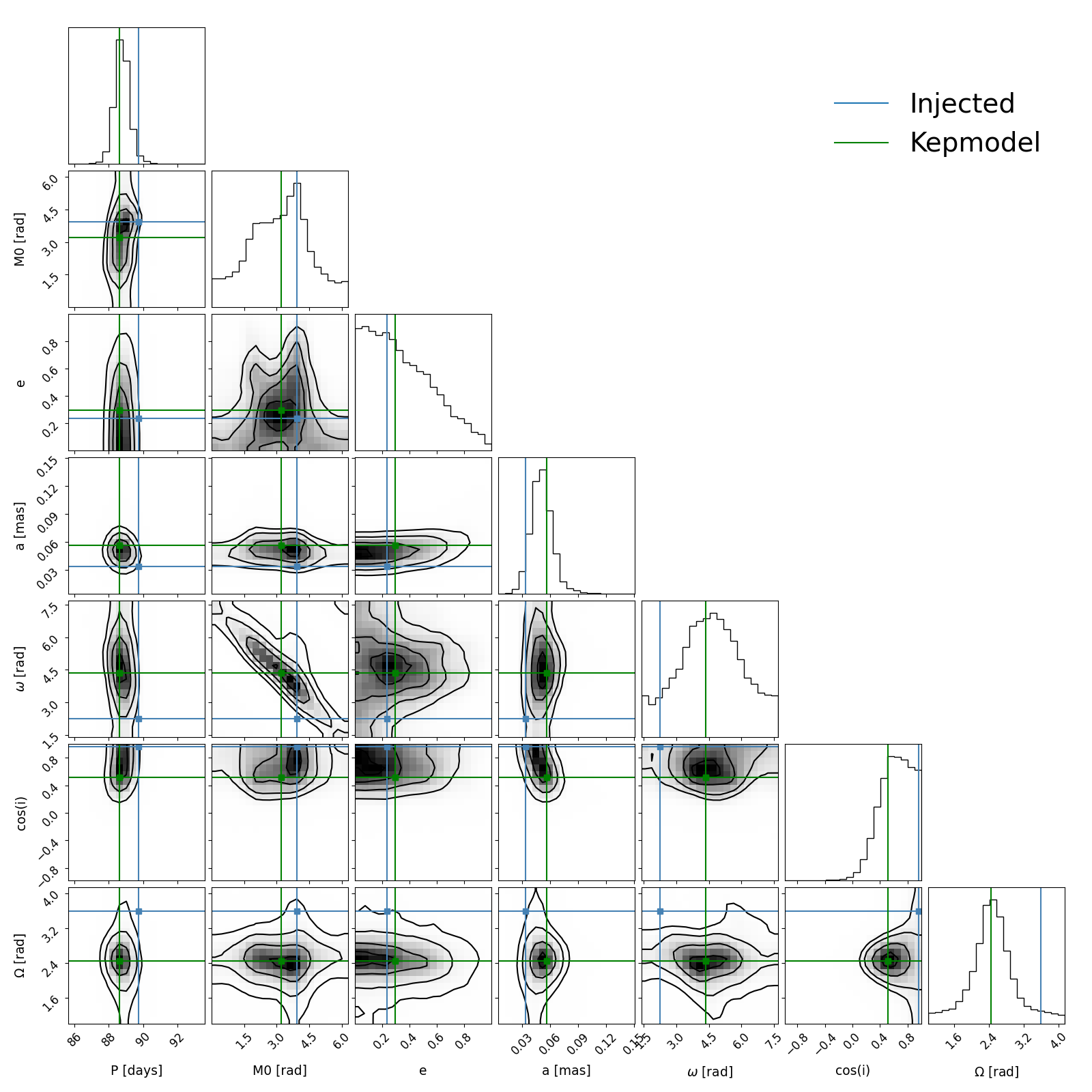}
    \caption{Corner plot of the orbital parameters for Target 6 with the injected values shown.}
    \label{fig:Target6_corner}
\end{figure}

\begin{figure}
    \centering
    \includegraphics[width=0.9\columnwidth]{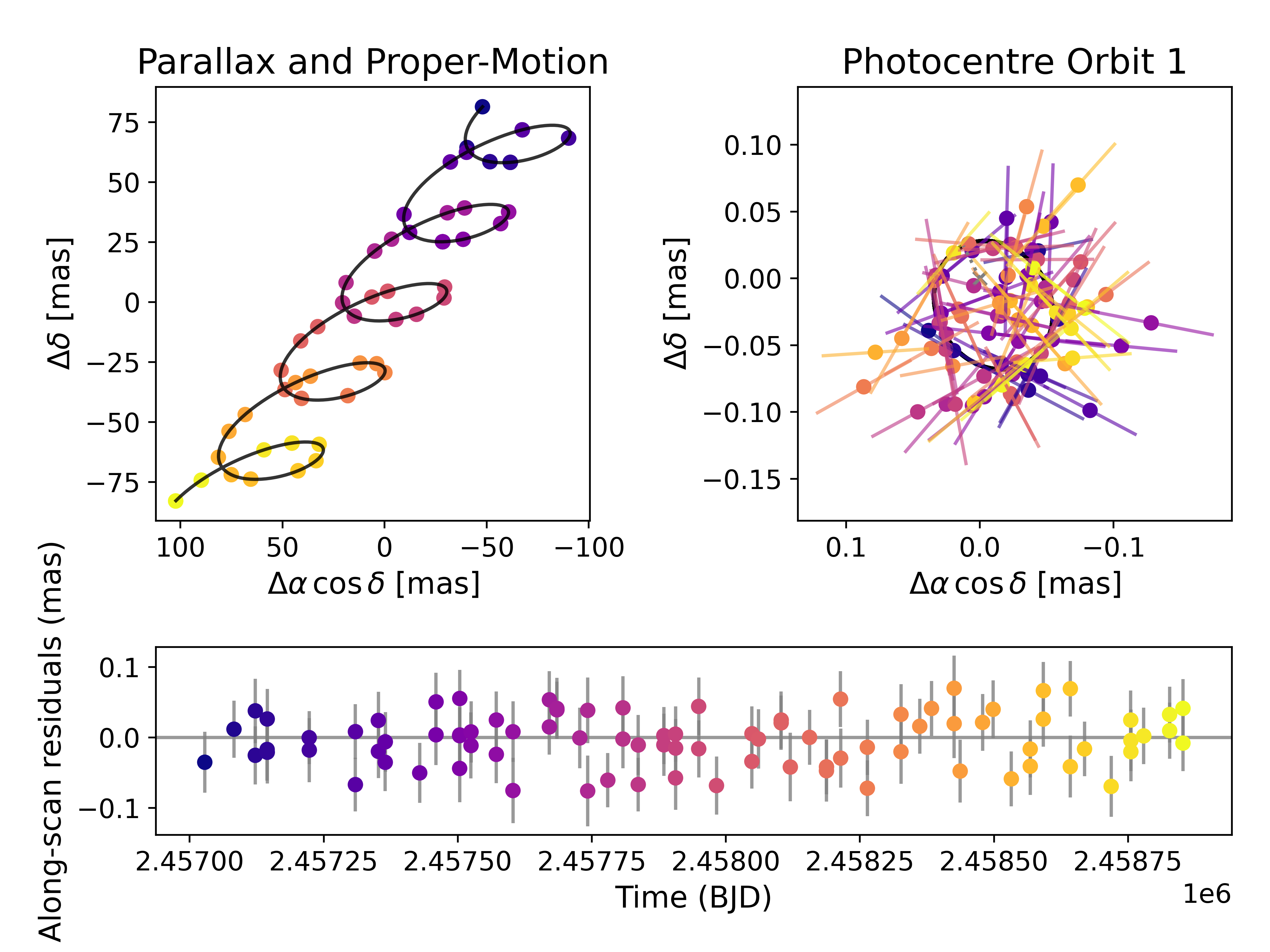}
    \caption{Phase plot of the maximum-likelihood model for Target 6, constrained to \( \pi \leq \Omega \leq 4\). Top-left: Parallax and proper-motion with orbit signal removed. Top-right: Orbit of the photocentre around the centre-of-mass. Bottom: residuals in the along-scan direction. The colour corresponds to the time.}
    \label{fig:Target6_phase_correct}
\end{figure}

\begin{figure}
    \centering
    \includegraphics[width=\linewidth]{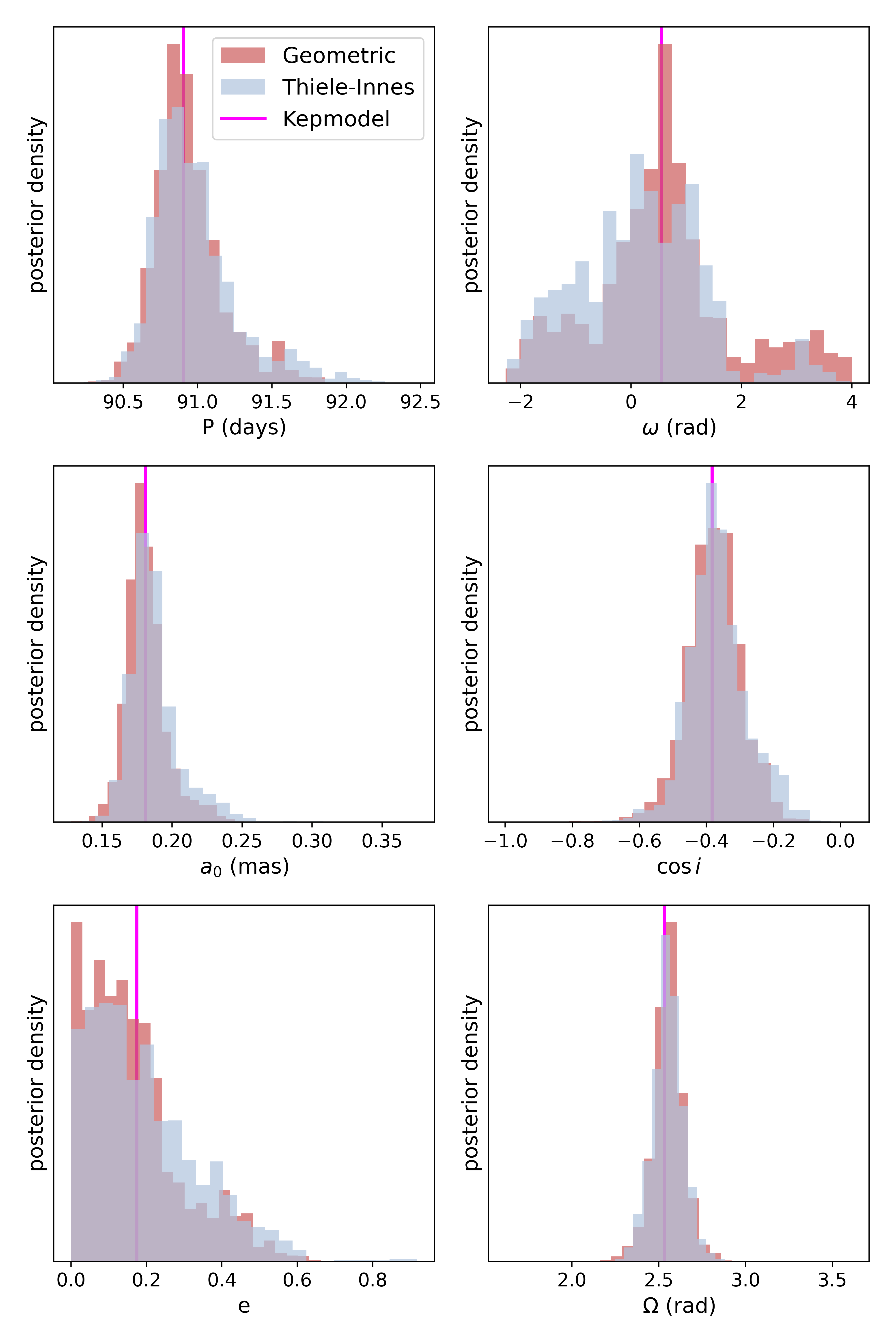}
    \caption{Posterior distributions for \(P,a,e,\omega,\cos{i},{\rm \,and\,}\Omega\) for analysis for Target 5 comparing results from fitting with geometric and Thiele-Innes parameterisations. \edit{The likelihood-maximised value from \kepmodel is also shown.}}
    \label{fig:param_comp_TI}
\end{figure}

\begin{figure}
    \centering
    \includegraphics[width=0.9\columnwidth]{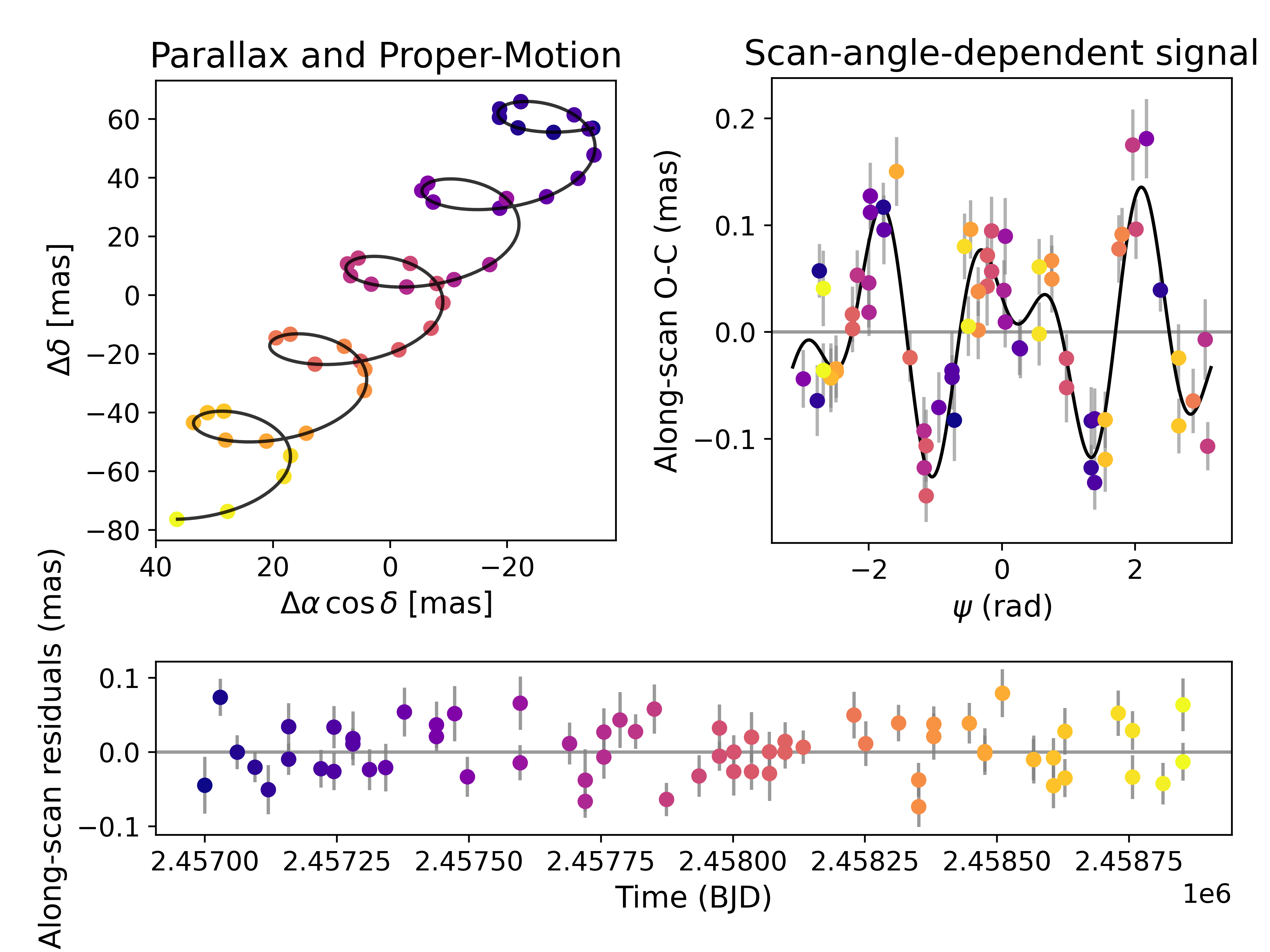}
    \caption{Phase plot of the maximum-likelihood model for Target 4. Top-left: Parallax and proper-motion with scan-angle dependent signal removed. Top-right: Two-harmonic signal as a function of scan-angle. Bottom: residuals in the along-scan direction. The colour corresponds to the time.}
    \label{fig:Target4_phase}
\end{figure}

\begin{figure}
    \centering
    \includegraphics[width=0.9\columnwidth]{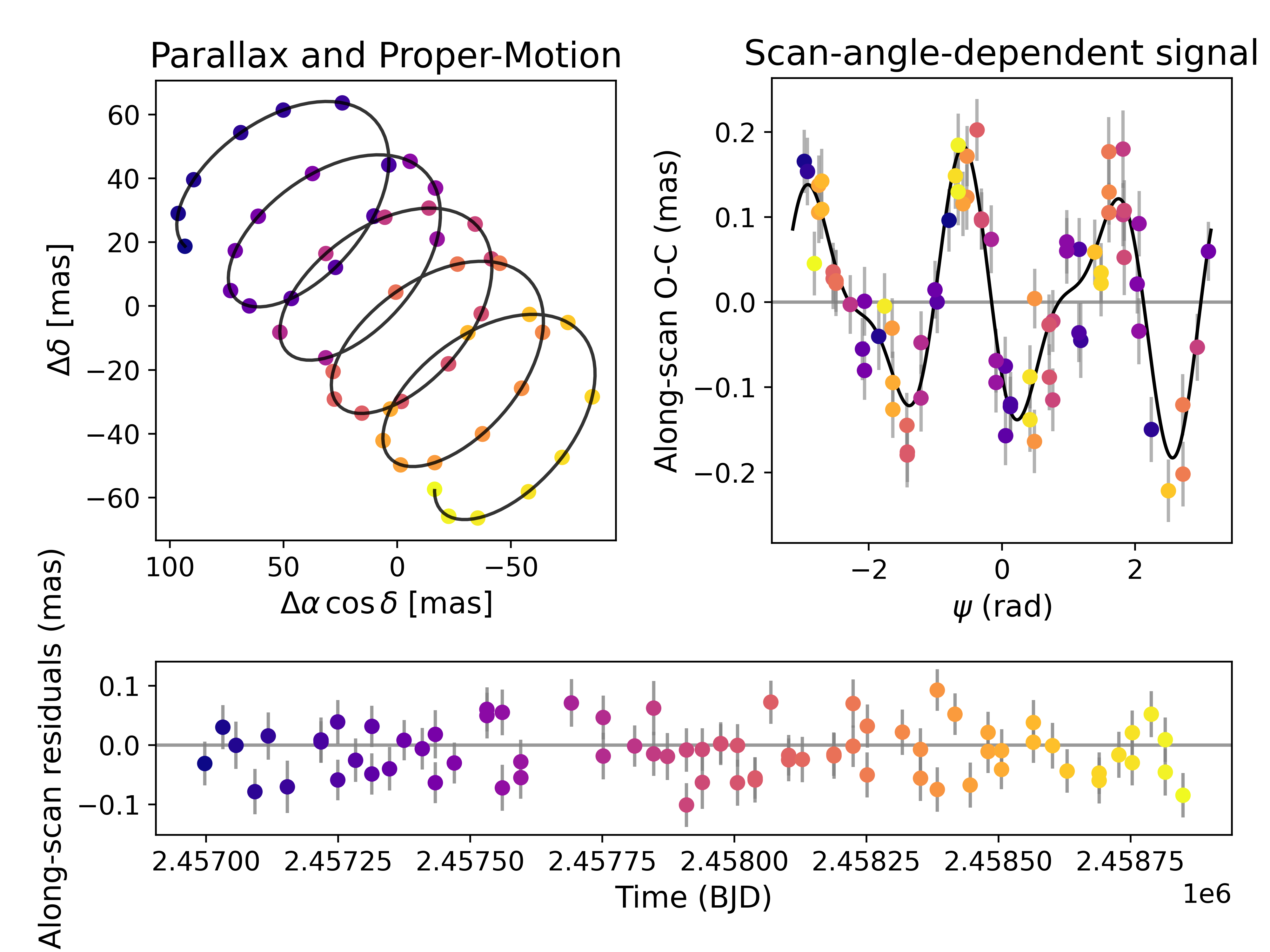}
    \caption{Phase plot of the maximum-likelihood model for Target 5. Top-left: Parallax and proper-motion with scan-angle dependent signal removed. Top-right: Two-harmonic signal as a function of scan-angle. Bottom: residuals in the along-scan direction. The colour corresponds to the time.}
    \label{fig:Target5_phase_sa}
\end{figure}

\begin{figure}
    \centering
    \includegraphics[width=0.9\columnwidth]{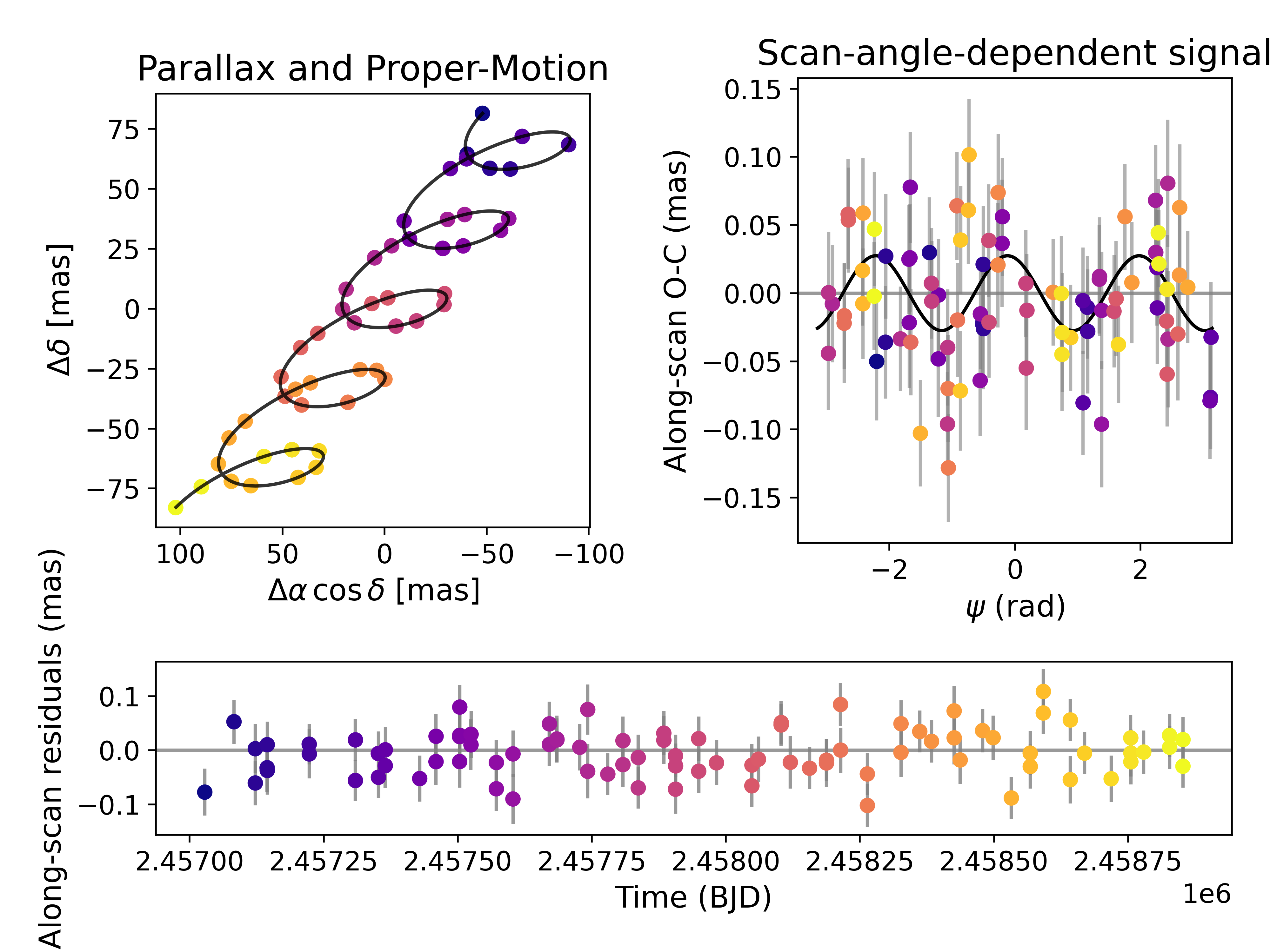}
    \caption{Phase plot of the maximum-likelihood model for Target 6. Top-left: Parallax and proper-motion with scan-angle dependent signal removed. Top-right: One-harmonic signal as a function of scan-angle. Bottom: residuals in the along-scan direction. The colour corresponds to the time.}
    \label{fig:Target6_phase_sa}
\end{figure}

\begin{figure}
    \centering
    \includegraphics[width=0.9\columnwidth]{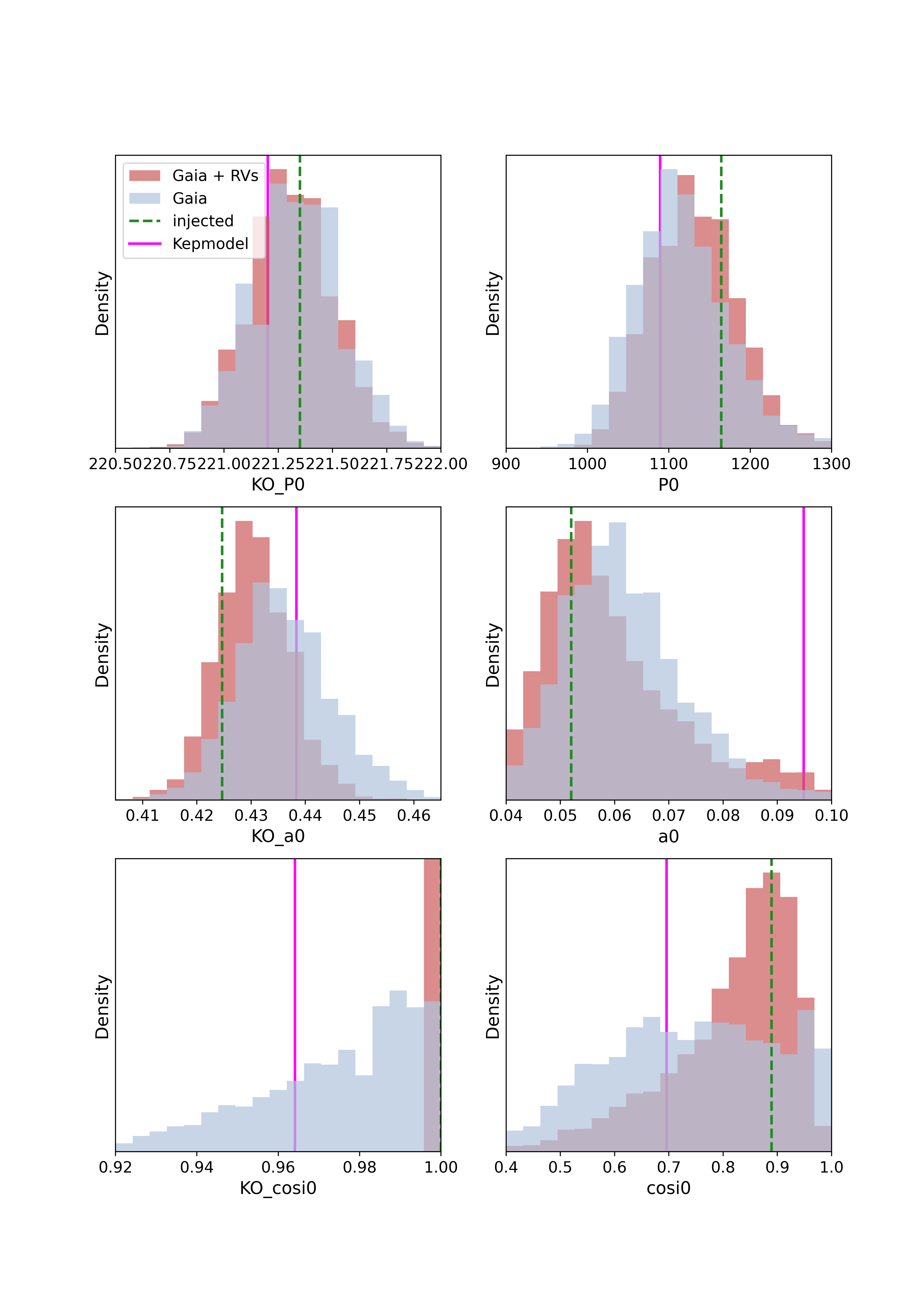}
    \caption{\edit{Posterior distributions for the analyses of the two-Keplerian injection into the Target 2 data. The results from analysing the \gaia data alone as well as jointly with the simulated radial velocities are shown, alongside the injected value and the optimized parameters from \kepmodel.}}
    \label{fig:multibody}
\end{figure}

\begin{figure}
    \centering
    \includegraphics[width=0.9\columnwidth]{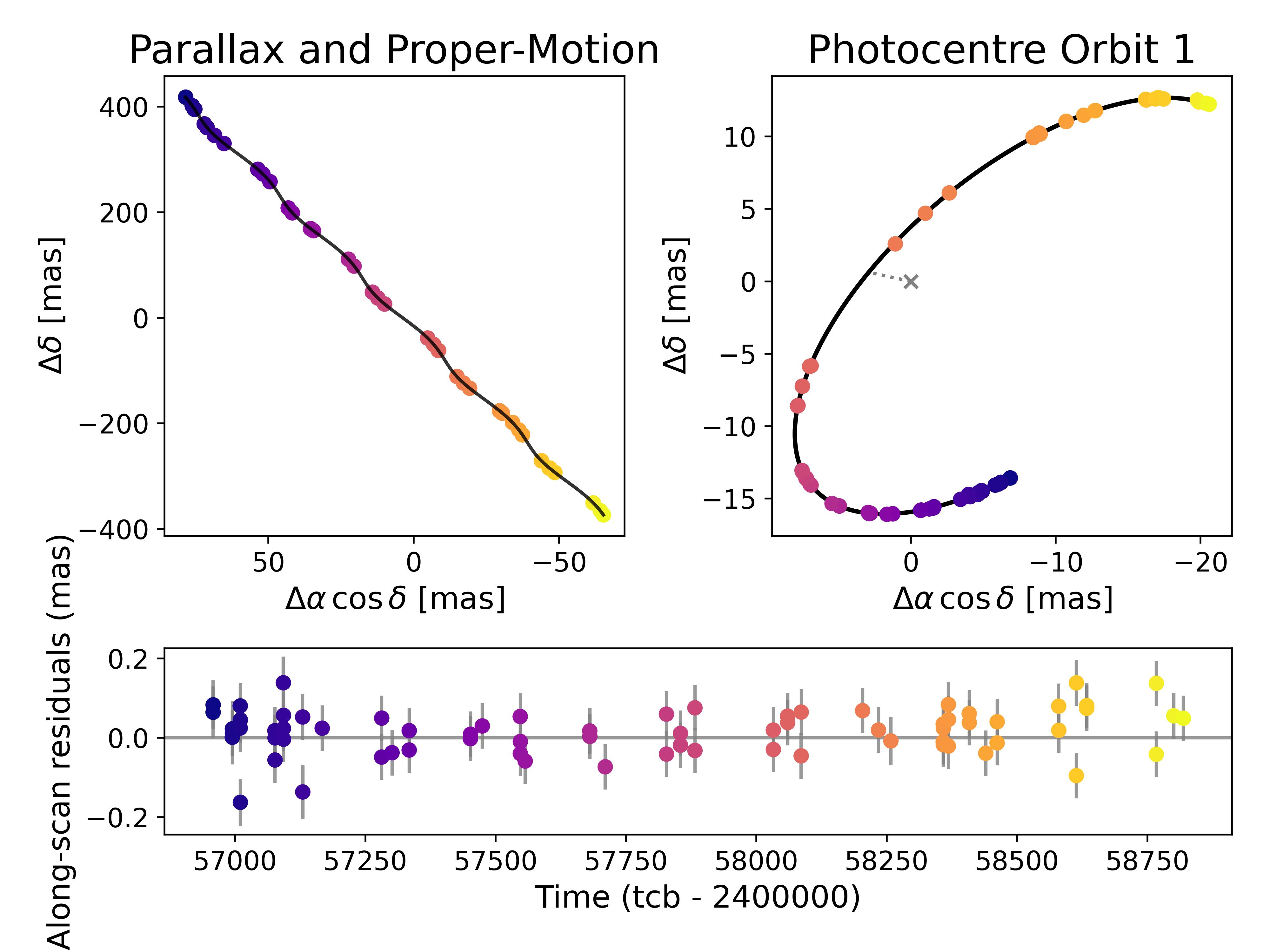}
    \caption{Phase plot of the maximum-likelihood model for \gaia BH3 from the GAIAmodel fit. Top-left: Parallax and proper-motion with orbit signal removed. Top-right: Orbit of the photocentre around the centre-of-mass. Bottom: residuals in the along-scan direction. The colour corresponds to the time.}
    \label{fig:BH3_GAIA_phase}
\end{figure}

\begin{figure}
    \centering
    \includegraphics[width=0.9\columnwidth]{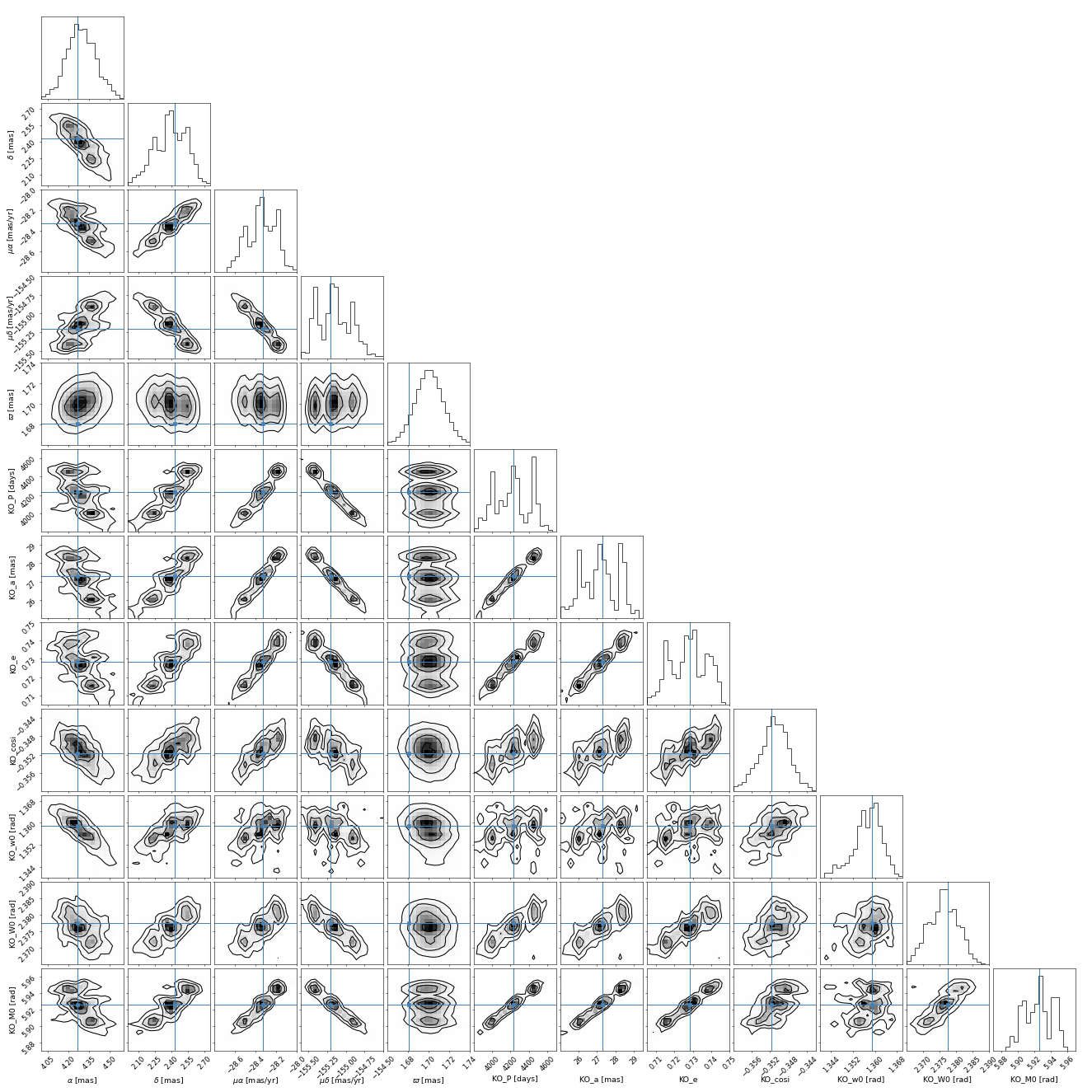}
    \caption{Corner plot of the five-parameter astrometric parameters and the orbital parameters for \gaia BH3 from the GAIAmodel fit, with the values obtained using \kepmodel shown.}
    \label{fig:BH3_GAIA_corner}
\end{figure}

\begin{figure}
    \centering
    \includegraphics[width=0.9\columnwidth]{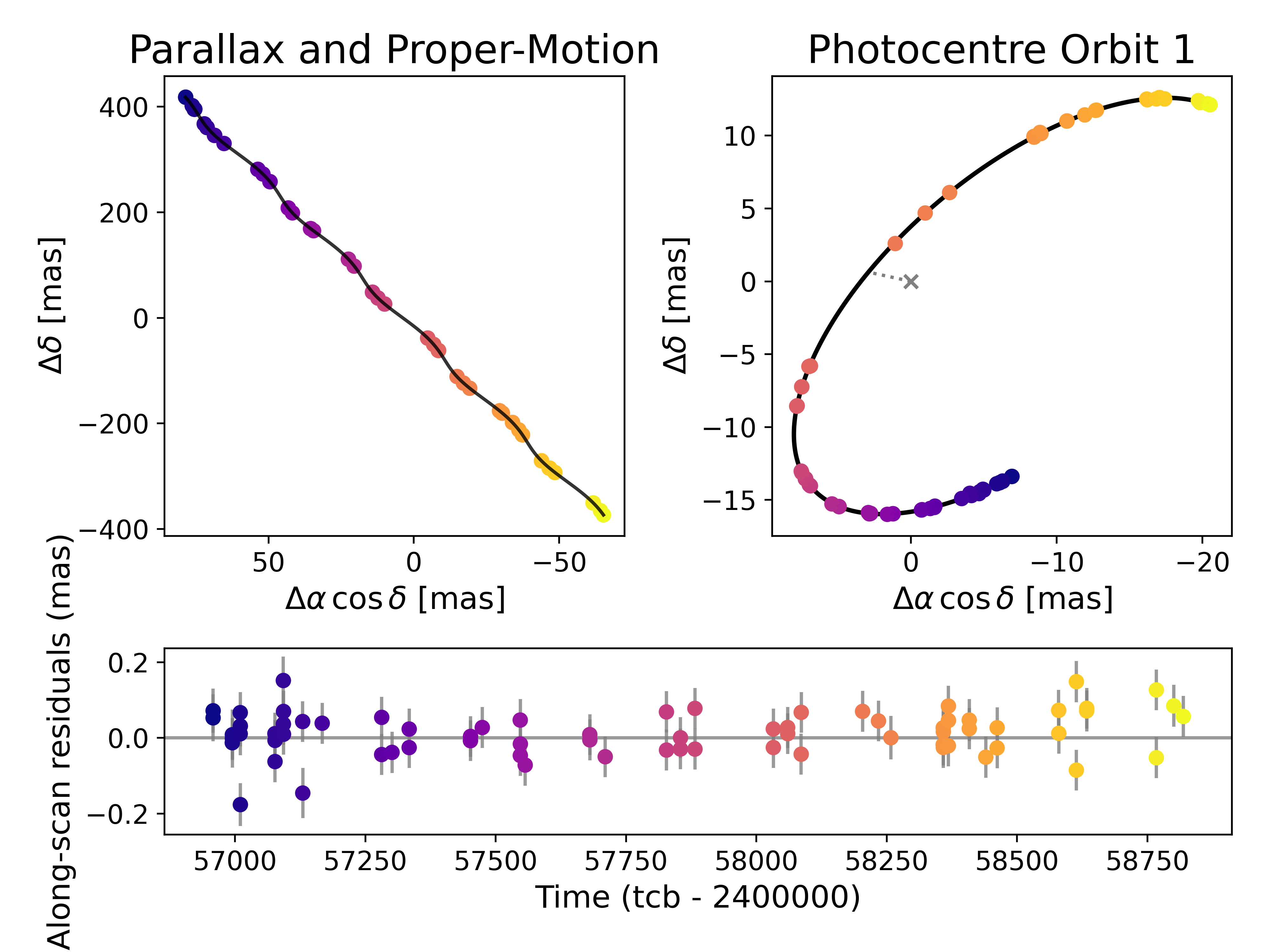}
    \caption{Astrometric phase plot of the maximum-likelihood model for \gaia BH3 from the RVGAIAmodel fit. Top-left: Parallax and proper-motion with orbit signal removed. Top-right: Orbit of the photocentre around the centre-of-mass. Bottom: residuals in the along-scan direction. The colour corresponds to the time.}
    \label{fig:BH3_RVGAIA_phase_ast}
\end{figure}

\begin{figure}
    \centering
    \includegraphics[width=0.9\columnwidth]{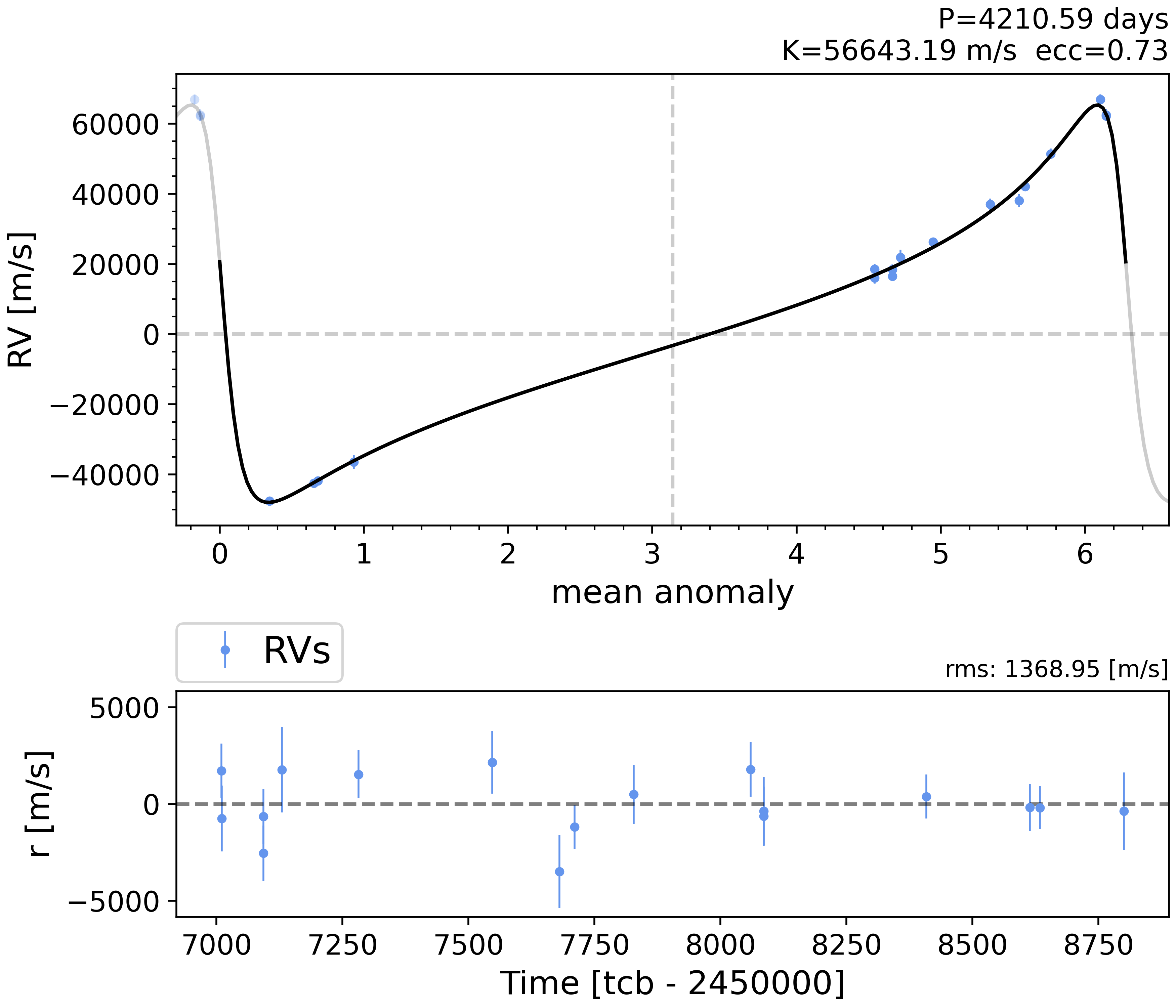}
    \caption{Radial velocity phase plot of the maximum-likelihood model for \gaia BH3 from the RVGAIAmodel fit}
    \label{fig:BH3_RVGAIA_phase_rv}
\end{figure}

\begin{figure}
    \centering
    \includegraphics[width=0.9\columnwidth]{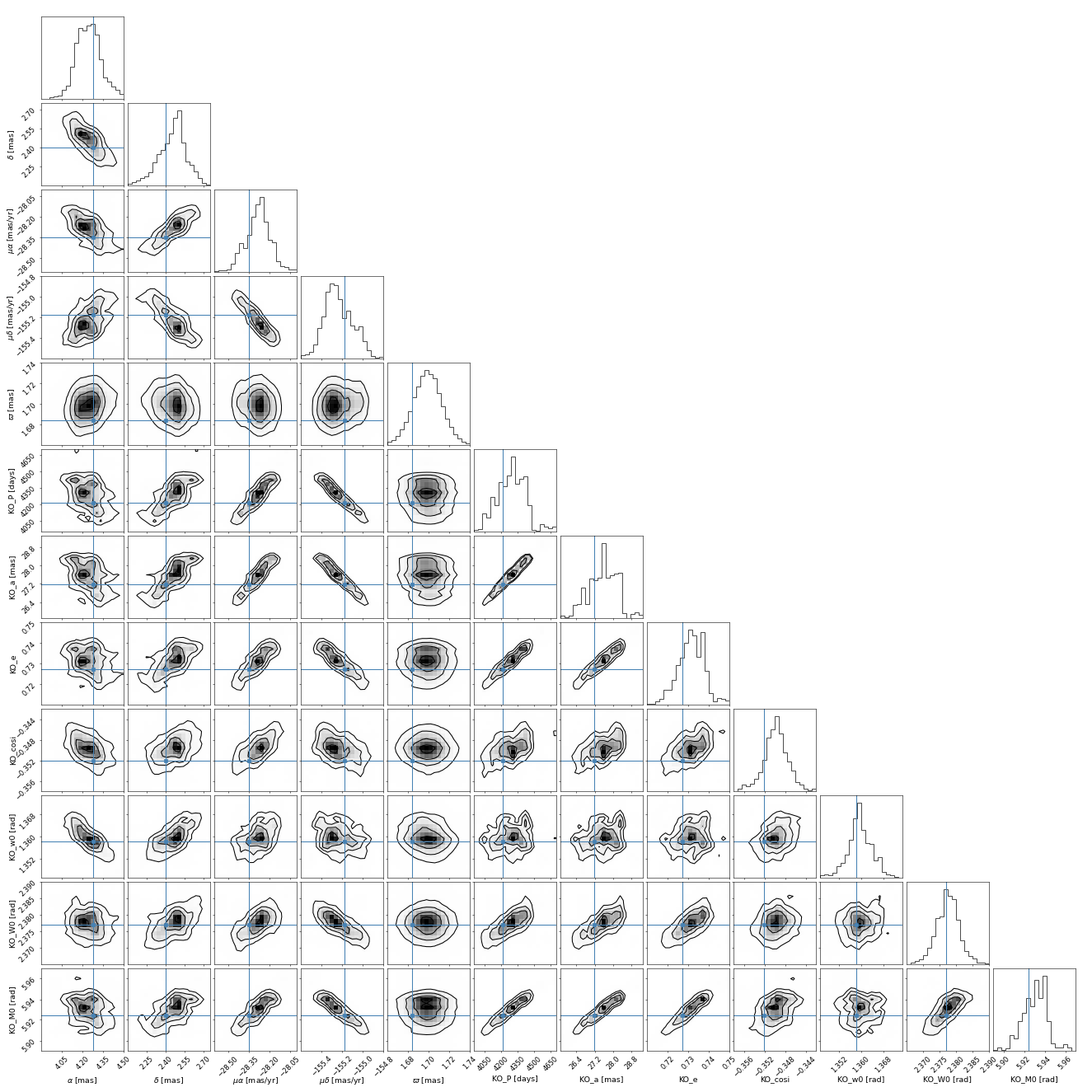}
    \caption{Corner plot of the five-parameter astrometric parameters and the orbital parameters for \gaia BH3 from the RVGAIAmodel fit, with the values obtained using {\tt kepmodel} shown.}
    \label{fig:BH3_RVGAIA_corner}
\end{figure}

\begin{figure}
    \centering
    \includegraphics[width=\linewidth]{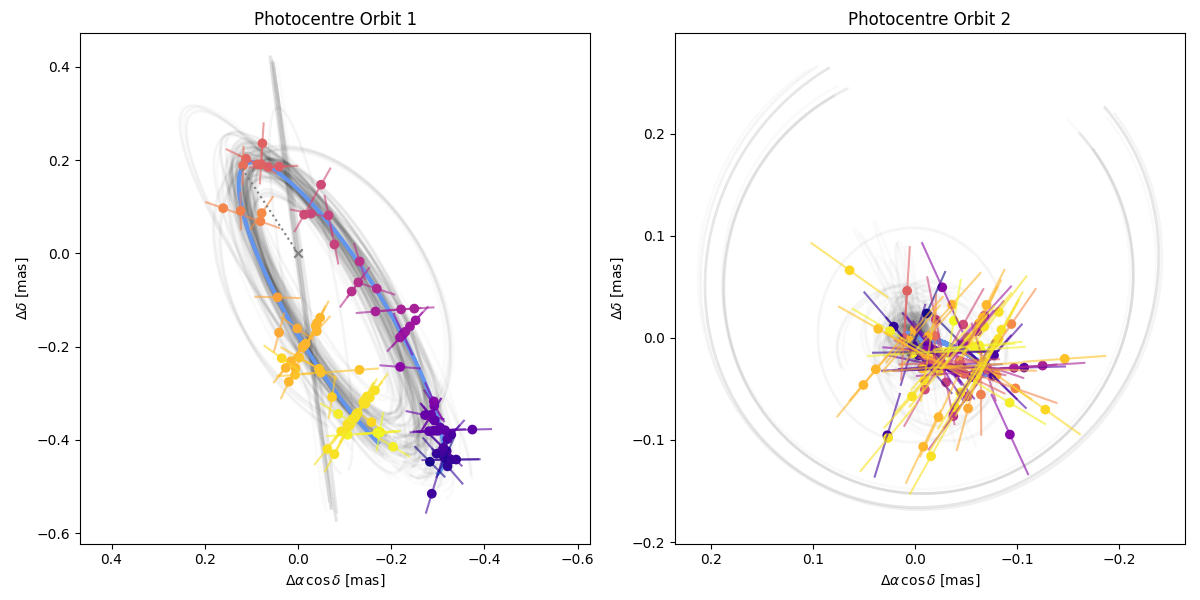}
    \caption{Data and model from the maximum-likelihood solution for 2 Keplerians. 200 random draws from the 2-Keplerian posterior samples are also shown. Note that these cannot be directly compared to the values of the data shown since the values of the baseline astrometry is different.}
    \label{fig:2plsamps}
\end{figure}

\end{appendix}

\end{document}